\pdfoutput=1
\documentclass[sn-mathphys-num]{sn-jnl}

\usepackage{graphicx}%
\usepackage{multirow}%
\usepackage{amsmath,amssymb,amsfonts}%
\usepackage{amsthm}%
\usepackage{mathrsfs}%
\usepackage{array}
\usepackage[title]{appendix}%
\usepackage{xcolor}%
\usepackage{textcomp}%
\usepackage{manyfoot}%
\usepackage{booktabs}%
\usepackage{algorithm}%
\usepackage{algorithmicx}%
\usepackage{algpseudocode}%
\usepackage{listings}%
\usepackage{lineno,hyperref}    
\usepackage{caption}
\usepackage{subfigure} 
\usepackage{upgreek} 
\usepackage{tabularx}
\usepackage{amsmath,amsthm,amssymb,amsfonts}
\usepackage{siunitx}
\usepackage{threeparttable}
\usepackage{multirow}
\usepackage{makecell}
\modulolinenumbers[5]   

\theoremstyle{thmstyleone}%
\theoremstyle{thmstyletwo}%

\theoremstyle{thmstylethree}%

\begin{document}

\title[Article Title]{QiandaoEar22: A high quality noise dataset for identifying specific ship from multiple underwater acoustic targets using ship-radiated noise}


\author[1,2]{\fnm{Xiaoyang} \sur{Du}}

\author*[1]{\fnm{Feng} \sur{Hong}}\email{hongfeng@mail.ioa.ac.cn}

\affil[1]{\orgname{Shanghai Acoustics Laboratory, Chinese Academy of Sciences}, \orgaddress{\street{Jiading District}, \city{Shanghai}, \postcode{201815}, \country{China}}}

\affil[2]{\orgdiv{University of Chinese Academy of Sciences},  \orgaddress{\street{Haidian District}, \city{Beijing}, \postcode{100190}, \country{China}}}


\abstract{Target identification of ship-radiated noise is a crucial area in underwater target recognition. However, there is currently a lack of multi-target ship datasets that accurately represent real-world underwater acoustic conditions. To tackle this issue, we conducted experimental data acquisition, resulting in the release of QiandaoEar22 \textemdash a comprehensive underwater acoustic multi-target dataset. This dataset encompasses 9 hours and 28 minutes of real-world ship-radiated noise data and 21 hours and 58 minutes of background noise data. To demonstrate the availability of QiandaoEar22, we executed two experimental tasks. The first task focuses on assessing the presence of ship-radiated noise, while the second task involves identifying specific ships within the recognized targets in the multi-ship mixed data. In the latter task, we extracted eight features from the data and employed six deep learning networks for classification, aiming to evaluate and compare the performance of various features and networks. The experimental results reveal that ship-radiated noise can be successfully identified from background noise in over 99\% of cases. Additionally, for the specific identification of individual ships, the optimal recognition accuracy achieves 99.56\%. Finally, based on our findings, we provide advice on selecting appropriate features and deep learning networks, which may offer valuable insights for related research.  Our work not only establishes a benchmark for algorithm evaluation but also inspires the development of innovative methods to enhance UATD and UATR systems.}




\maketitle

\section{Introduction}\label{sec1}

The significance of underwater target sound as a vital source of information has been underscored since the invention of sonar by American Fessenden in 1913 \cite{d2009brief}. Consequently, underwater acoustic technology has been progressively applied, facilitating the differentiation of various sounding target types and enabling analysis of environmental signals. 
Notably, the focus has shifted towards the ship-radiated noise due to its enhanced concealment, and security.
\begin{figure*}[t!]  
	
	\begin{center}
		\centering
		\subfigure[SpeedBoat]{
			\includegraphics[width=0.28\textwidth,height=1in]{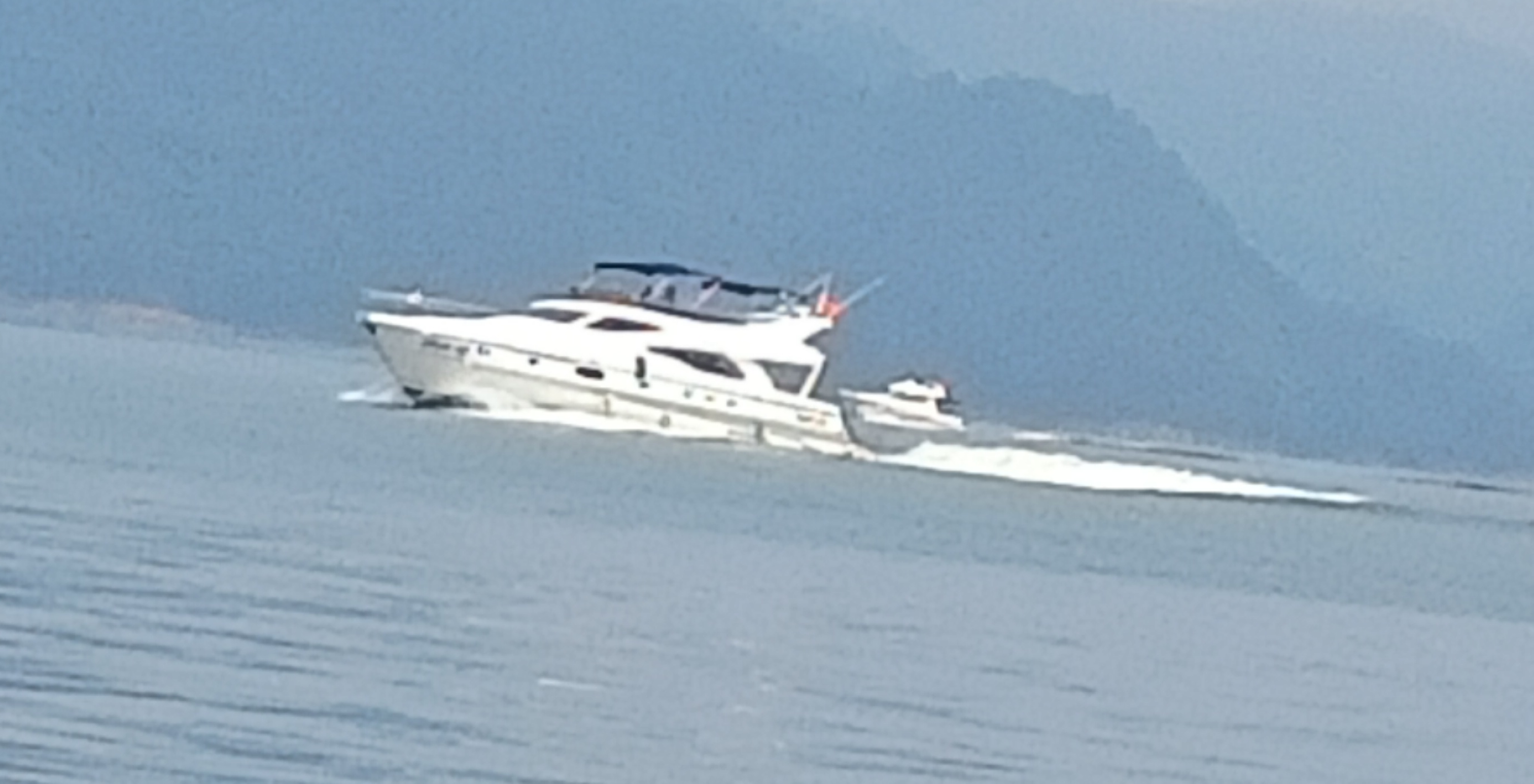}
		}
		\subfigure[KaiYuan]{
			\includegraphics[width=0.28\textwidth,height=1in]{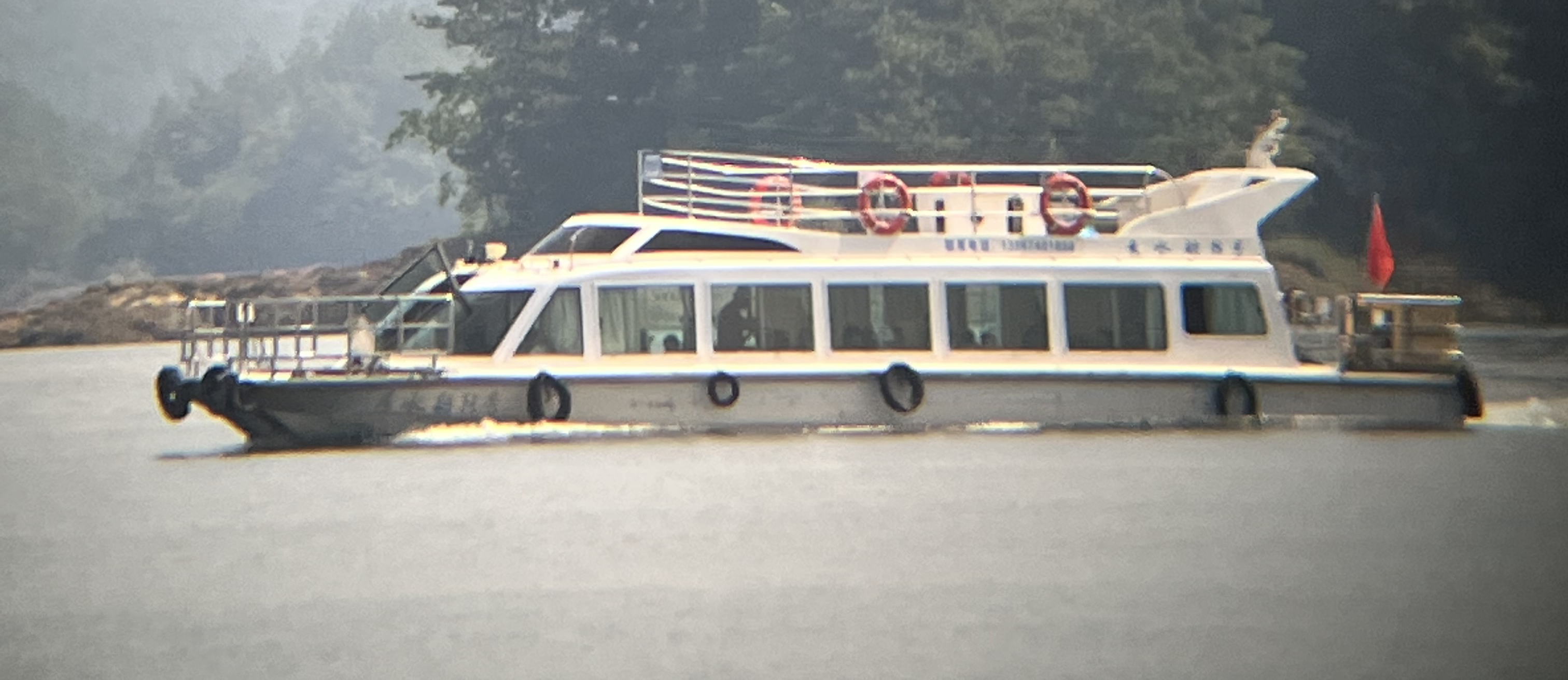}
		}
		\subfigure[UUV]{
			\includegraphics[width=0.28\textwidth,height=1in]{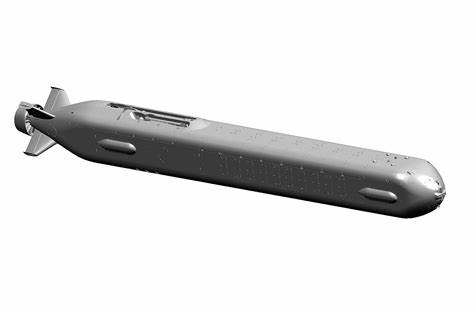}
		}
		\caption{Target ships}
		\label{boat}
	\end{center}
\end{figure*}

However, publicly available datasets for underwater acoustic target research are scarce for two primary reasons. Firstly, military secrecy restricts access to numerous recordings made in military facilities \cite{zak2008ships}. Secondly, the high cost associated with conducting underwater experiments, compounded by signal loss during propagation and limited data quality due to underwater communication equipment constraints, presents challenges in dataset creation \cite{luo2023survey,neupane2020review}. Data, labeling further contribute to the time and energy consumption.

Some small scale datasets have been created, including those from cargo ships, navy vessels, and personal watercraft \cite{zak2008ships,arveson2000radiated,lennartsson2006fused,mckenna2012underwater,erbe2013underwater,roth2013underwater}. While some recent datasets like ShipsEar \cite{santos2016shipsear} and Deepship \cite{irfan2021deepship} are available, they primarily consist of single-target ship data in simple environments, lacking the complexity of multi-target scenarios often encountered in real-world conditions.

To address this issue, we present the QiandaoEar22 dataset, containng 21 hours and 58 minutes of background noise data and 9 hours and 28 minutes of multi-target ship noise signal data. Collected using digitalHyd SR-1 self-capacitance hydrophones in a real underwater acoustic environment at Qiandao Lake, China, during June 2022, this dataset serves as a high-quality ship-radiated noise recording. 

Underwater acoustic target identification focuses on discerning the class of ship-radiated noise from background noise. Additionally, underwater acoustic target detection aims to determine the presence or absence of ship targets through binary classification, classifying received signals as either ship-radiated or background signals. This preliminary classification streamlines subsequent tasks. To accomplish this, we first identify ship-radiated noise and background noise signals from acquired underwater acoustic data. Extracting three 1D features (spectrum, power spectral density, DEMON spectra \cite{chung2011demon}) and five 2D features (LOFAR spectrum \cite{chen2021underwater}, Log Mel spectrum, MFCC \cite{zhang2016feature}, PNCC \cite{kim2016power}, GFCC \cite{zhao2013analyzing}) for the multi-target ship sound signal. Then we employ diverse deep neural networks, including CNN \cite{Cheng0Research}, CRNN \cite{fu2017crnn}, BiLSTM \cite{Zhang2020Classification},DenseNet \cite{yu2021additive}, ResNet \cite{he2016deep}, and ECAPA-TDNN \cite{desplanques2020ecapa} to classify these features. Finally three specific types of ship signals: speedboat, KaiYuan and UUV are identified from the multi-target ship signals, respectively, as shown in Fig.\ref{boat}. Furthermore, we explore the labeling and formatting aspects of the dataset. Our objective is that the dataset, along with the experimental results, assists researchers in benchmarking algorithms and promotes the development of innovative methods to improve Underwater Acoustic Target Detection (UATD) and Underwater Acoustic Target Recognition (UATR) systems.

The main contribution of this paper is as follows.

Firstly, we have experimentally collected and constructed a high-quality underwater acoustic dataset named QiandaoEar22. This dataset comprises 21 hours and 58 minutes of background noise data and 9 hours and 28 minutes of multi-target vessel noise signal data. QiandaoEar22 dataset closely simulates real recognition environments, providing researchers with a valuable resource to comprehensively evaluate and enhance their underwater acoustic target recognition algorithms.

Secondly, using QiandaoEar22 dataset, this article first distinguishes ship signals and background noise signals from the underwater acoustic data. Then we successfully recognize three specific types of ship signals \textemdash speedboats, KaiYuan, and UUV \textemdash from the multi-target ship signals, showcasing the practicality of our dataset.

Lastly, we simplifies the identification of specific ship signals from multi-target ship signals by extracting three spectral features and five t-f features. These features are tested using six popular deep learning classifiers, offering researchers a benchmark for algorithm evaluation and some advice on selecting appropriate features and deep learning networks, which may offer valuable insights for related research.

Section \ref{sec2} details the experiments on acquisition and labeling of the QiandaoEar22 dataset. Section \ref{sec3} introduces the evaluation of experimental tasks, encompassing the experimental setup, presentation of the two tasks, and analysis and discussion of the results. Conclusions are presented in Section \ref{sec4}.

\section{The QiandaoEar22 dataset}\label{sec2}
\subsection{Recording system and methodology}\label{subsec1}

\begin{figure*}[t!]  
	\centering
	\includegraphics[height=3in]{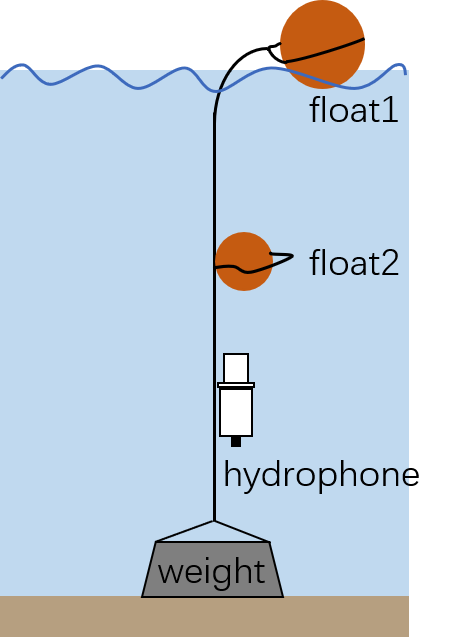}
	\caption{ Schematic diagram of the experimental hydrophone to be deployed in Qiandao Lake in June 2022}
	\label{exp}
\end{figure*}

The data collection took place from June 24 to 28, 2022, at the Xin' anjiang Experimental Site of the Institute of Acoustics, Chinese Academy of Sciences (CAS), situated in the town of Qiandao Lake, Zhejiang, China. Two primary deployment sites were utilized, with coordinates at 118.974532\textdegree E, 29.557795\textdegree N, and 118.947097\textdegree E, 29.548708\textdegree N, respectively. Throughout the deployment, a hydrophone was anchored to the center of the deployment area using a counterweight and buoy system, as depicted in Fig.\ref{exp}. The water depth ranged from approximately 30 to 50 meters, and the hydrophone was positioned at a depth of 10 to 15 meters. A dedicated hydrophone was employed for the data collection, using a stand-alone setup positioned near the channel to capture a broader range of radiated noise from the ships.

\begin{figure*}[t!]  
	\centering
	\includegraphics[width=0.8\textwidth,height=0.8in]{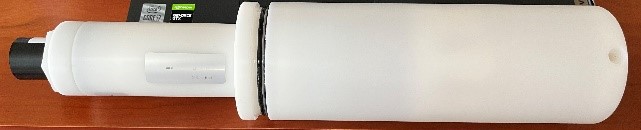}
	\caption{DigitalHyd SR-1 hydrophone}
	\label{hydro}
\end{figure*}

The hydrophone model used is DigitalHyd SR-1, illustrated in Fig. \ref{hydro}, with key specifications provided in Table \ref{speci}. The DigitalHyd SR-1 features a sample frequency of 52,734 Hz, a sample resolution of 16 bits, a usable acoustic band of 1~25.8 kHz, a receive sensitivity of -162.2$\sim$-126.1 dB re 1V/$\upmu $Pa, and an input sound pressure level range of 46.3 dB re 1$\upmu$ Pa$\sim$172.5 dB re 1$\upmu $ Pa. The operating depth can reach up to 100 meters, constructed with plastic steel, weighing approximately 0.77 kg. It operates within a temperature range of 0$\sim$\SI{40}{\degreeCelsius} and employs a USB interface. The device is powered by a rechargeable lithium-ion battery, enabling continuous data collection for up to 12 hours. Acoustic data is stored in WAV format on a removable memory card, which also retains all configuration parameters for subsequent data analysis.

\begin{table*}[t!]
	\centering
	\caption{DigitalHyd SR-1 Hydrophone specifications}
	{\fontsize{8}{10}\selectfont	
	\label{speci}
	\resizebox{\linewidth}{!}{
		\begin{tabular}{ll}
			\Xhline{1pt}
			Index & Parameter \\
			\Xhline{0.5pt}
			Sample frequency & 52,734 Hz\\
			Sample resolution & 16 bits\\
			Usable Acoustic Band & 1$\sim$25.8kHz \\
			Receive sensitivity & -162.2$\sim$-126.1dB re 1V/$\upmu$Pa\\
			Input sound pressure level range & 46.3dB re 1$\upmu$Pa$\sim$172.5dB re 1$\upmu$Pa\\
			Operation depth & 100m\\
			Surface material & plastic steel\\
			Weight & 0.77kg\\
			Operation temperature & 0$\sim$\SI{40}{\degreeCelsius}\\
			Device interface & USB interface\\
			Power supply & rechargeable lithium-ion battery\\
			\Xhline{1pt}
		\end{tabular}
	}
	}
\end{table*}

\subsection{Dataset structure}\label{subsec2}

\begin{figure*}[htbp]  
	\centering
	\includegraphics[width=1\textwidth]{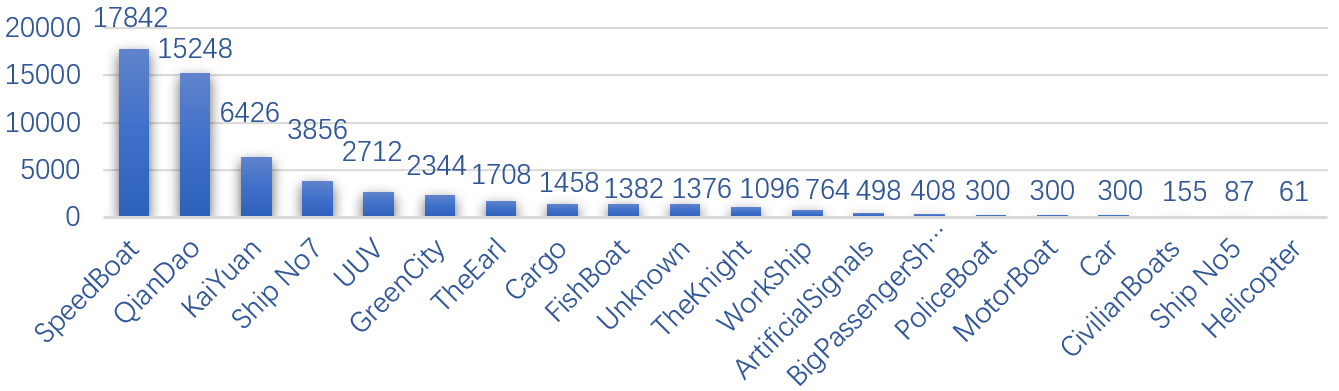}
	\caption{Duration of sound data for all vessels}
	\label{distribution}
\end{figure*}

Following the hydrophone deployment, we observed ships navigating through the channel. We recorded the time when a ship approached and moved away from the hydrophone, approximately 1 kilometer away, along with the ship number and name at those moments. Subsequent data alignment involves identifying the moment at the waveform peak and associating it with the corresponding ship number, thereby correlating the labeling and data for that specific segment.The experimental period coincided with the tourist season, resulting in recorded ship-radiated noise primarily originating from sightseeing boats and speedboats. Due to the hydrophone's proximity to the navigation channel, passing vessels, including sightseeing boats, speedboats, and other watercraft, contribute to a composite radiated noise. Additional targets include large tour boats, cargo ships, sanitation vessels, small boats, research ships, and ambient noise.

Based on experimentally recorded document information and signal amplitude and spectral characteristics, we determine the start and end time nodes for each audio segment. We label each data piece based on auditory sensation, distance, and labeling data, recording the information in a text document. During this process, the data is segmented into 3-second intervals for ease of subsequent processing. The final database consists of 10,611 records in WAV format, totaling 9 hours, 28 minutes, and 13 seconds of ship-radiated noise data. Each audio file, along with its corresponding data, forms a record in the QiandaoEar22 dataset. The example information fields for each record is as following:

20220624102345 \_ DATA0008 \_ M \_ GreenCity \_ N \_ S \& SpeedBoat \_ M \_ W

(start/stop time \_ single/multiple targets(S/M) \_ ship name \_ distance(N/M/F) \_ audibility(S/M/W) \& name \_ distance \_ audibility)
\begin{itemize}
	\item[$\bullet$]"filename": filename.
	\item[$\bullet$]"time": The start and end time of the audio recording.
	\item[$\bullet$]"single/multiple target":The audio capture object is a single ship target or multiple ship targets, which is denoted by S for single target and M for multiple targets.
	\item[$\bullet$]"name": name of the vessel.
	\item[$\bullet$]"distance": Indicates the distance of the vessel from the hydrophone, N for near, M for medium and F for far.
	\item[$\bullet$]"audibility": indicates the audibility of the voice, S for strong, M for Middle, W for weak.
	\item[$\bullet$]If multi-target, link with \& in order, for each ship including "name", "distance", "audibility ".
\end{itemize}

\begin{table*}[t!]\scriptsize
	\centering
	\caption{Background-Target ship Dataset}
	{\fontsize{8}{10}\selectfont	
	\label{bttarget}
	\centering
		\begin{tabular}{p{4.04em}|cc|cc|cc|cc}
			\Xhline{1pt}
			\multicolumn{1}{c|}{Dataset} & \multicolumn{2}{c|}{Train} & \multicolumn{2}{c|}{Test} & \multicolumn{2}{c|}{Valid} & \multicolumn{2}{c}{Total} \\
			\Xhline{0.5pt}
			\multicolumn{1}{c|}{Kind}  & \multicolumn{1}{c|}{target} & \multicolumn{1}{c|}{back} & \multicolumn{1}{c|}{target} & \multicolumn{1}{c|}{back} & \multicolumn{1}{c|}{target} & \multicolumn{1}{c|}{back} & \multicolumn{1}{c|}{target} & \multicolumn{1}{c}{back} \\
			\Xhline{0.5pt}
			\multicolumn{1}{c|}{Num}   & \multicolumn{1}{c|}{3500} & 3550  & \multicolumn{1}{c|}{750} & 700   & \multicolumn{1}{c|}{750} & 750   & \multicolumn{1}{c|}{5000} & 5000 \\
			\Xhline{0.5pt}
			\multicolumn{1}{c|}{Time(s)} & \multicolumn{1}{c|}{10500} & 10650 & \multicolumn{1}{c|}{2250} & 2100  & \multicolumn{1}{c|}{2250} & 2250  & \multicolumn{1}{c|}{15000} & 15000 \\
			\Xhline{0.5pt}
			\multicolumn{1}{c|}{Total time(s)} & \multicolumn{2}{c|}{21150} & \multicolumn{2}{c|}{2350} & \multicolumn{2}{c|}{4500} & \multicolumn{2}{c}{30000}\\
			\Xhline{1pt}
		\end{tabular}%
	}
\end{table*}

Based on the consistency of labels, we identified a total of 143 recordings containing both single and multiple ship targets. These data primarily include 20 categories of ship targets, and the overall time distribution of data containing individual targets is illustrated in Fig. \ref{distribution}.

Simultaneously, we selected time slices without targets and other interference to constitute the background noise data, labeled as files DATA0XXX. Ultimately, we obtained a total of 21 hours and 58 minutes, comprising 25,900 records of background noise data, each with a duration of 3 seconds. The dataset can be download following the instructions on https://github.com/xiaoyangdu22/QiandaoEar22. 

\section{Ship noise detection experimental tasks and analysis of results}\label{sec3}
\label{3}

Underwater acoustic target detection focuses on discerning the presence or absence of a target and classifying the received signal as either underwater target radiated noise or non-target noise, involving binary classification. Hence, using this dataset, we undertook two tasks. Task 1 involves identifying ship signals from the acquired underwater acoustic signal data, which includes background noise signals. In Task 2, our objective is to identify three specific types of ship signals \textemdash speedboats, KaiYuan, and UUVs \textemdash respectively, from the mixed multi-target ship-radiated noise signals.

\subsection{Task 1: determine whether the audio contains ship-radiated noise}\label{subsec3}

The objective of Task 1 is to determine whether the audio contains ship-radiated noise. We ultimately selected the first 5000 ship noise recordings, along with the initial 30 recordings from DATA0001 and 20 recordings from DATA0002, to compose the final dataset. Data segmentation details has been introduced in Part 2. We use 70\% of the dataset for training, 15\% for validation, and 15\% for test. The characteristics of the final dataset are presented in Table \ref{bttarget}. Log Mel spectra features were extracted for the data. Deep learning networks were employed for feature classification, and the network structures used are shown in Table \ref{networkst}. Each experiment was conducted ten times to calculate the mean and variance.

Table \ref{acc1} presents the accuracy results for all deep learning methods. Excep BiLSTM, the classification accuracy for other methods exceeded 99\%. Notably, the CRNN exhibited the most robust classification performance, achieving an accuracy of 99.79\%. Additionally, its recall, precision, F1 score, FNR, and FPR values also reached 99.79\%, 99.80\%, 99.79\%, 0.00\%, and 0.43\%, respectively. The experimental results indicate a high level of distinguishability between ship-radiated noise data and background noise data within the dataset, showcasing the significant application value of our dataset.

\begin{table*}[t!]\scriptsize
	\centering
	\caption{The structure of deep learning network}
	{\fontsize{8}{10}\selectfont
	\label{networkst}
	\centering
	\begin{threeparttable}
		\begin{tabular}{p{4.04em}p{32.00em}}
			\Xhline{1pt}
			Network  & \multicolumn{1}{c}{Structure} \\
			\Xhline{0.5pt}
			\multirow{9}{*}{CNN} & Conv\tnote{1} (cin=1, cout=4, k=5, s=2, p=1)+ BN(4) \\
			& + Dropout(0.1)+ ReLU+ MaxPool(k=2, s=2) \\
			& Conv(cin=4, cout=16, k=5, s=2, p=1)+ BN(16) \\
			& + Dropout(0.1)+ ReLU+ MaxPool(k=2, s=2) \\
			& Conv(cin=16, cout=32, k=3, s=2, p=1)+ BN(32) \\
			& + Dropout(0.1)+ ReLU+ MaxPool(k=1, s=1) \\
			& Conv(cin=32, cout=64, k=3, s=2, p=1)+ BN(64) \\
			& + Dropout(0.1)+ ReLU+ MaxPool(k=1, s=1) \\
			& Linear(64*128, 4*128)+ Linear(4*128, 32)+ Linear(32, 2) \\
			\Xhline{0.5pt}
			\multirow{10}{*}{CRNN} & Conv(cin =1, cout =4, k=5, s=2,p=1)+ BN(4)\\
				& + Dropout(0.1)+ ReLU+ MaxPool(k=2, s=2)\\
			 & Conv(cin =4, cout =16, k=5, s=2,p=1)+ BN(16)\\
			 & + Dropout(0.1)+ ReLU+ MaxPool(k=2, s=2)\\
			 & Conv(cin =16, cout =32, k=3, s=2,p=1)+ BN(32)\\
			 & + Dropout(0.1)+ ReLU+ MaxPool(k=1, s=1)\\
		     & Conv(cin =32, cout =64, k=3, s=2,p=1)+ BN(64)\\
		     & +Dropout(0.1)+ ReLU+ MaxPool(k=1, s=1) \\
			 & LSTM( 256,512, 2) \\
			 & Linear(512, 32) + Linear(32, 2) \\
			\Xhline{0.5pt}
			BiLSTM & LSTM( input, 256, 2) \\
			\Xhline{0.5pt}
			ResNet & ResNet18 \cite{he2016deep}( num classes=2) \\
			\Xhline{0.5pt}
			DenseNet & DenseNet121 \cite{yu2021additive} \\
			\Xhline{0.5pt}
			ECAPA-TDNN & \multirow{2}{*}{ECAPA-TDNN \cite{desplanques2020ecapa}} \\
			\Xhline{1pt}
		\end{tabular}%
		\begin{tablenotes}
			\footnotesize 
			\item[1] Note: Conv1d is used for 1D features and Conv2d for 2D features
		\end{tablenotes}
	\end{threeparttable}
}
\end{table*}

\begin{table*}[t!]\scriptsize
	\centering
	\caption{Accuracy Comparison (\%) for task 1}
	{\fontsize{8}{10}\selectfont
	\label{acc1}
	\begin{tabular}{>{\centering\arraybackslash}p{1.2cm}cccccc}
		\Xhline{1pt}
		net   & accuracy   & recall & precision   & f1    & FNR  &FPR \\
		\Xhline{0.5pt}
		CNN2d&99.66$\pm$0.00&99.65$\pm$0.00&99.66$\pm$0.00 & 99.65$\pm$0.00 & 0.27$\pm$0.00 & 0.43$\pm$0.00 \\
		CRNN2d&99.79$\pm$0.05&99.79$\pm$0.09&99.80$\pm$0.18 & 99.79$\pm$0.05 & 0.00$\pm$0.13 & 0.43$\pm$0.00 \\
		DenseNet&99.72$\pm$0.04&99.72$\pm$0.06&99.73$\pm$0.06 & 99.72$\pm$.05 & 0.13$\pm$0.21 & 0.43$\pm$0.06 \\
		ECAPA-TDNN&\multirow{2}{*}{99.31$\pm$0.00}&\multirow{2}{*}{99.42$\pm$0.00}&\multirow{2}{*}{99.29$\pm$0.01} & \multirow{2}{*}{99.42$\pm$0.00} & \multirow{2}{*}{0.67$\pm$0.02 }&\multirow{2}{*}{ 0.71$\pm$0.01} \\
		BiLSTM&91.17$\pm$0.01&91.37$\pm$0.00&91.61$\pm$0.01 & 91.17$\pm$0.01 & 14.27$\pm$0.03 & 0.30$\pm$0.00 \\
		ResNet&99.52$\pm$0.01&99.50$\pm$0.01&99.53$\pm$0.01 & 99.52$\pm$0.01 & 0.13$\pm$0.03 & 0.86$\pm$0.01 \\
		\Xhline{1pt}
	\end{tabular}
	}
\end{table*}

\begin{table*}[t!]\scriptsize
	\centering
	\caption{The SpeedBoat, KaiYuan and UUV datasets}
	{\fontsize{8}{10}\selectfont
	\label{3data}
	\centering
	\begin{tabular}{c|p{4.04em}|cc|cc|cc}
		\Xhline{1pt}
		\multicolumn{1}{c|}{data} & \multicolumn{1}{c|}{} & \multicolumn{2}{c|}{Train} & \multicolumn{2}{c|}{Test} & \multicolumn{2}{c}{Valid} \\
		\Xhline{0.5pt}
		\multicolumn{1}{c|}{\multirow{4}{*}{SpeedBoat}} & \multicolumn{1}{c|}{Kind}  & \multicolumn{1}{c|}{SpeedBoat} & \multicolumn{1}{c|}{Other} & \multicolumn{1}{c|}{SpeedBoat} & \multicolumn{1}{c|}{Other} & \multicolumn{1}{c|}{SpeedBoat} & \multicolumn{1}{c}{Other} \\
		\cline{2-8} \multicolumn{1}{c|}{}  &  \multicolumn{1}{c|}{Num}   & \multicolumn{1}{c|}{4848} & 2578  & \multicolumn{1}{c|}{1041} & 554   & \multicolumn{1}{c|}{1038} & 552 \\
		\cline{2-8} \multicolumn{1}{c|}{} &  \multicolumn{1}{c|}{Time(s)} & \multicolumn{1}{c|}{14544} & 7734  & \multicolumn{1}{c|}{3123} & 1662  & \multicolumn{1}{c|}{3114} & 1656 \\
		\cline{2-8} \multicolumn{1}{c|}{}  &  \multicolumn{1}{c|}{Total time(s)} & \multicolumn{2}{c|}{22278}& \multicolumn{2}{c|}{4785} & \multicolumn{2}{c}{4770}\\
		\Xhline{0.5pt}
		\multicolumn{1}{c|}{\multirow{4}{*}{KaiYuan}} & \multicolumn{1}{c|}{Kind}   & \multicolumn{1}{c|}{Kaiyuan} & \multicolumn{1}{c|}{Other} & \multicolumn{1}{c|}{Kaiyuan} & \multicolumn{1}{c|}{Other} & \multicolumn{1}{c|}{Kaiyuan} & \multicolumn{1}{c}{Other} \\
		\cline{2-8} \multicolumn{1}{c|}{} & \multicolumn{1}{c|}{Num}   & \multicolumn{1}{c|}{1363} & 2801  & \multicolumn{1}{c|}{293} & 599   & \multicolumn{1}{c|}{291} & 600 \\
		\cline{2-8} \multicolumn{1}{c|}{}   &  \multicolumn{1}{c|}{Time(s)} & \multicolumn{1}{c|}{4089} & 8403  & \multicolumn{1}{c|}{879} & 1797  & \multicolumn{1}{c|}{873} & 1800 \\
		\cline{2-8} \multicolumn{1}{c|}{}  & \multicolumn{1}{c|}{Total time(s)} & \multicolumn{2}{c|}{12492} & \multicolumn{2}{c|}{2676} & \multicolumn{2}{c}{2673} \\
		\Xhline{0.5pt}
		\multicolumn{1}{c|}{\multirow{4}{*}{UUV}} & \multicolumn{1}{c|}{Kind}   & \multicolumn{1}{c|}{UUV} & \multicolumn{1}{c|}{Other} & \multicolumn{1}{c|}{UUV} & \multicolumn{1}{c|}{Other} & \multicolumn{1}{c|}{UUV} & \multicolumn{1}{c}{Other} \\
		\cline{2-8} \multicolumn{1}{c|}{} & \multicolumn{1}{c|}{Num}   & \multicolumn{1}{c|}{631} & 1401  & \multicolumn{1}{c|}{136} & 299   & \multicolumn{1}{c|}{135} & 300 \\
		\cline{2-8} \multicolumn{1}{c|}{} & \multicolumn{1}{c|}{Time(s)} & \multicolumn{1}{c|}{1893} & 4203  & \multicolumn{1}{c|}{408} & 897   & \multicolumn{1}{c|}{405} & 900 \\
		\cline{2-8} \multicolumn{1}{c|}{}& \multicolumn{1}{c|}{Total time(s)} & \multicolumn{2}{c|}{6096} & \multicolumn{2}{c|}{1305} & \multicolumn{2}{c}{1305}\\
		\Xhline{1pt}
	\end{tabular}%
	}		
\end{table*}
\begin{figure*}[t!]  
	\begin{center}
		\centering
		\subfigure[Frequency spectrum ( 16537 )]{
			\includegraphics[width=0.21\textwidth]{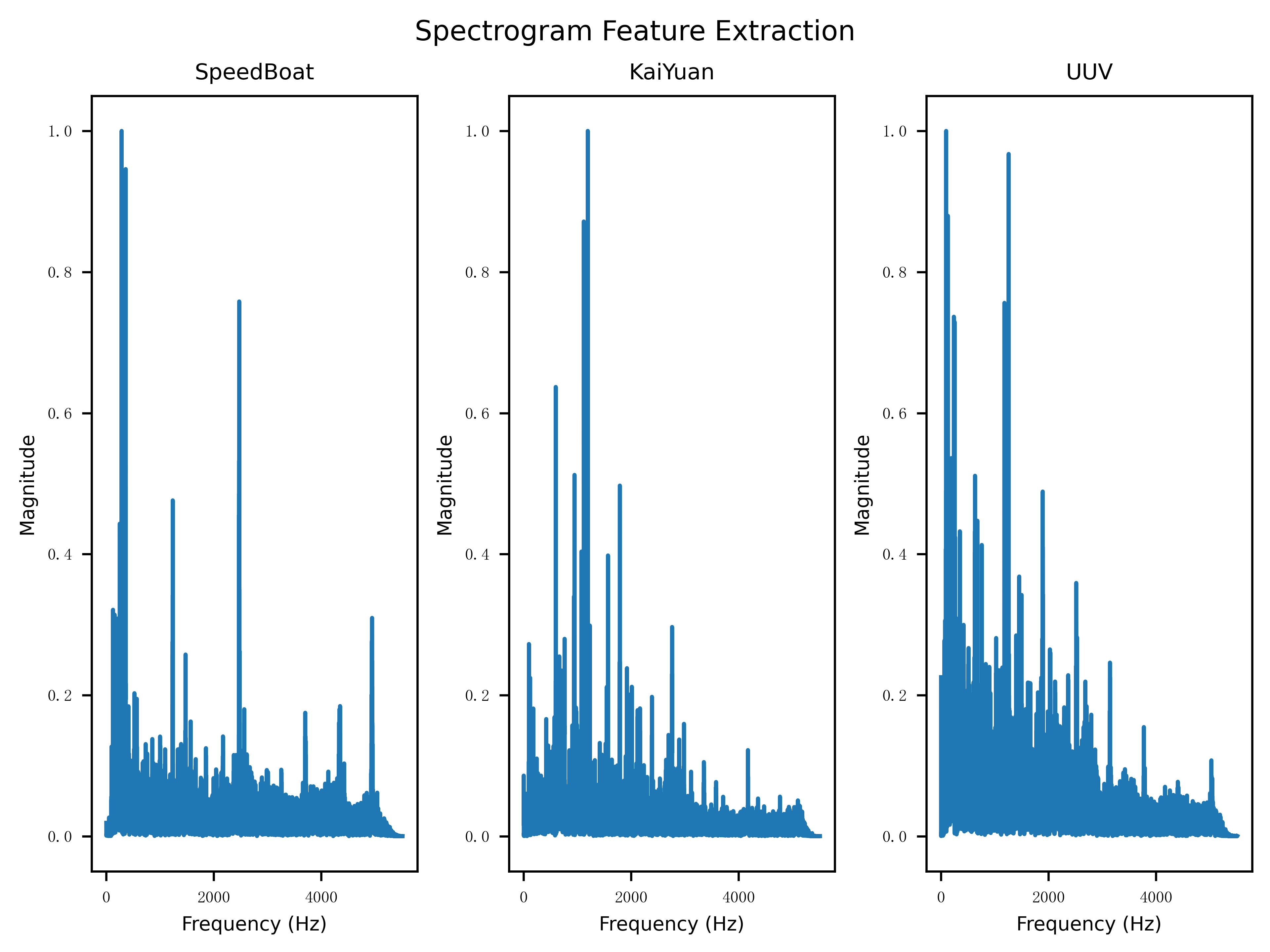}
		}
		\subfigure[PSD ( 16385 )]{
			\includegraphics[width=0.21\textwidth]{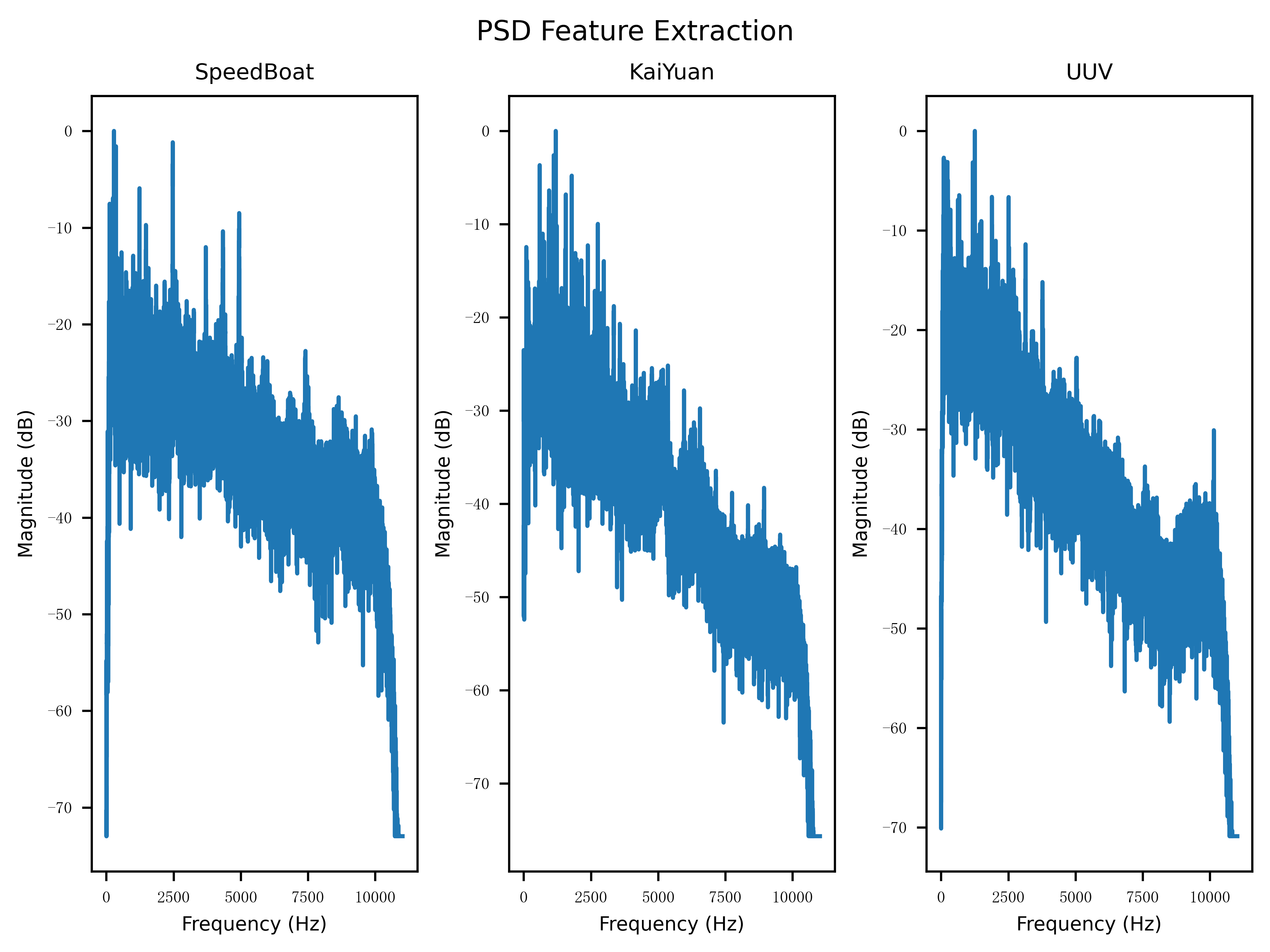}
		}
		\subfigure[DEMON ( 16384 )]{
			\includegraphics[width=0.21\textwidth]{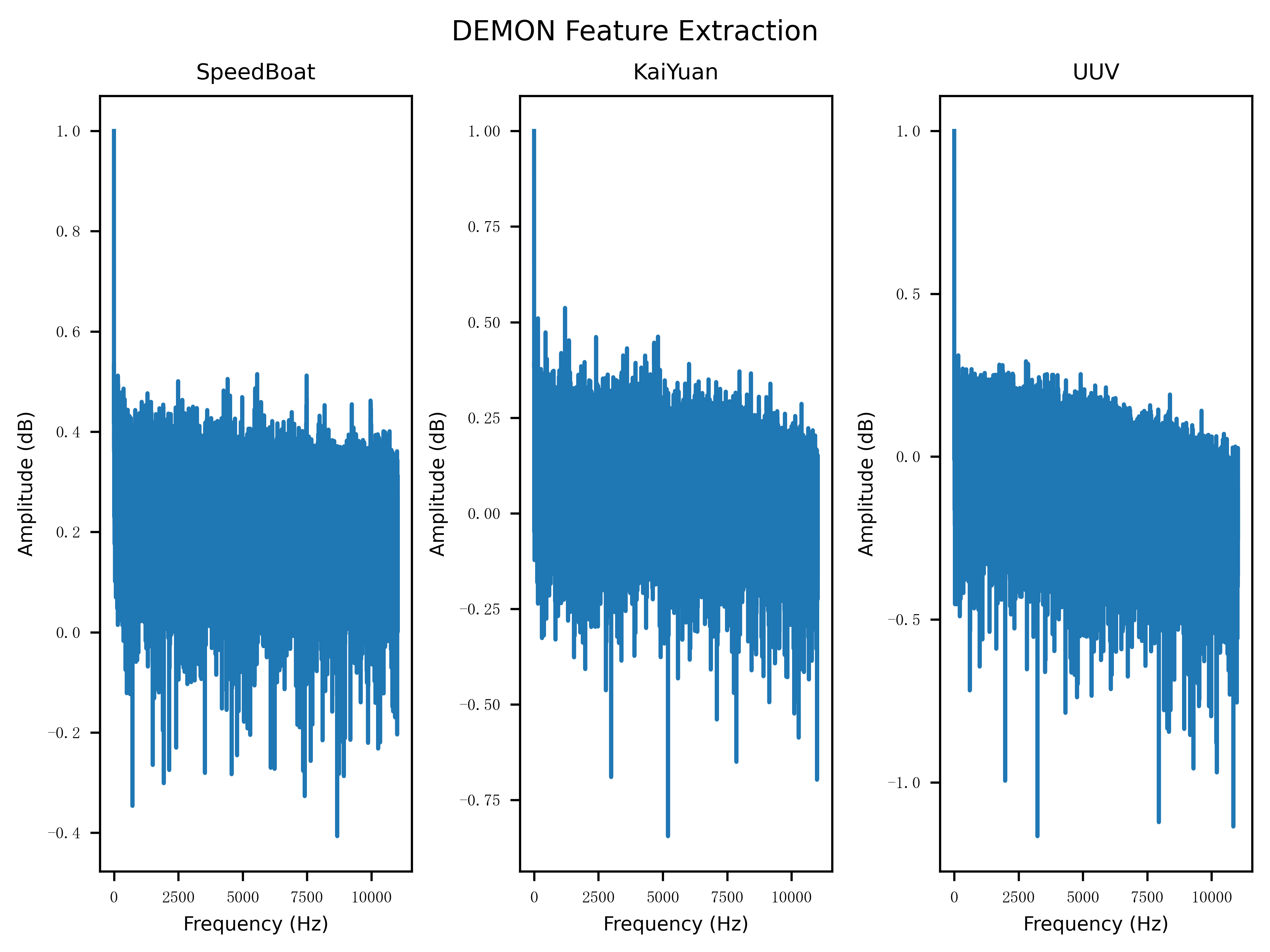}
		}
		\subfigure[LOFAR ( 129*131 = 16899 )]{
			\includegraphics[width=0.21\textwidth]{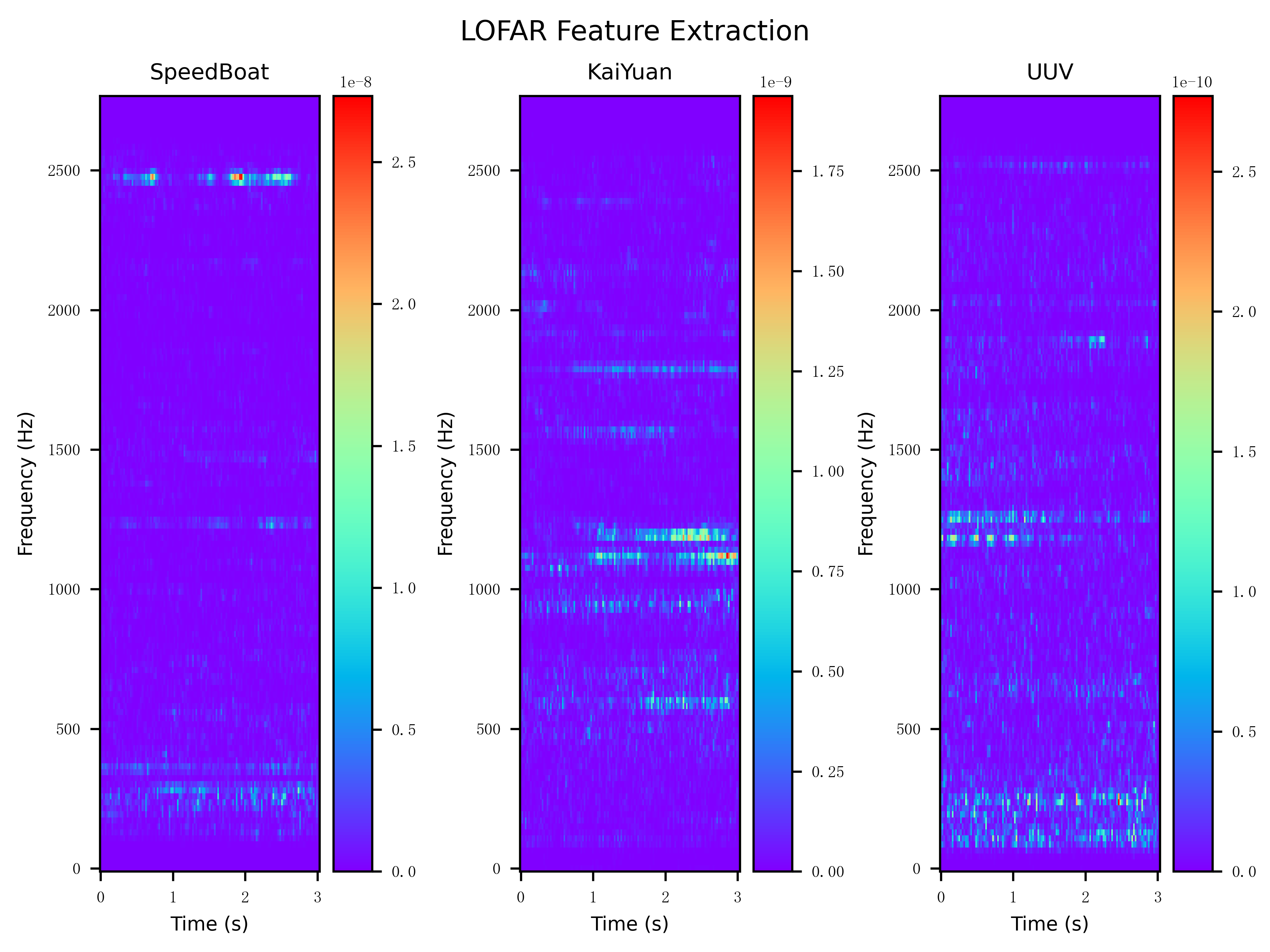}
		}
		
		\subfigure[Log Mel ( 128*130 = 16640 )]{
			\includegraphics[width=0.21\textwidth]{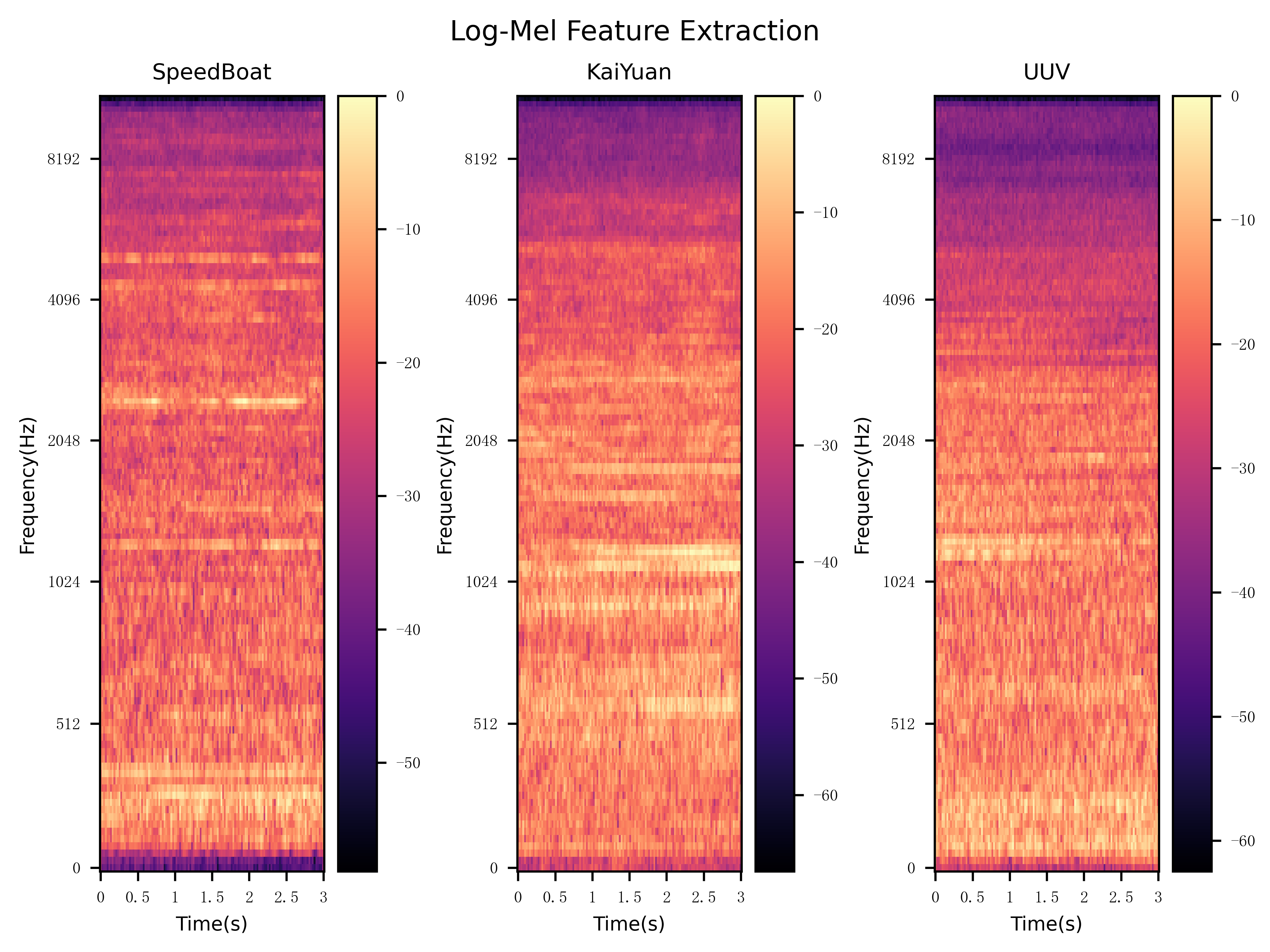}
		}
		\subfigure[MFCC ( 127*130 = 16510 )]{
			\includegraphics[width=0.21\textwidth]{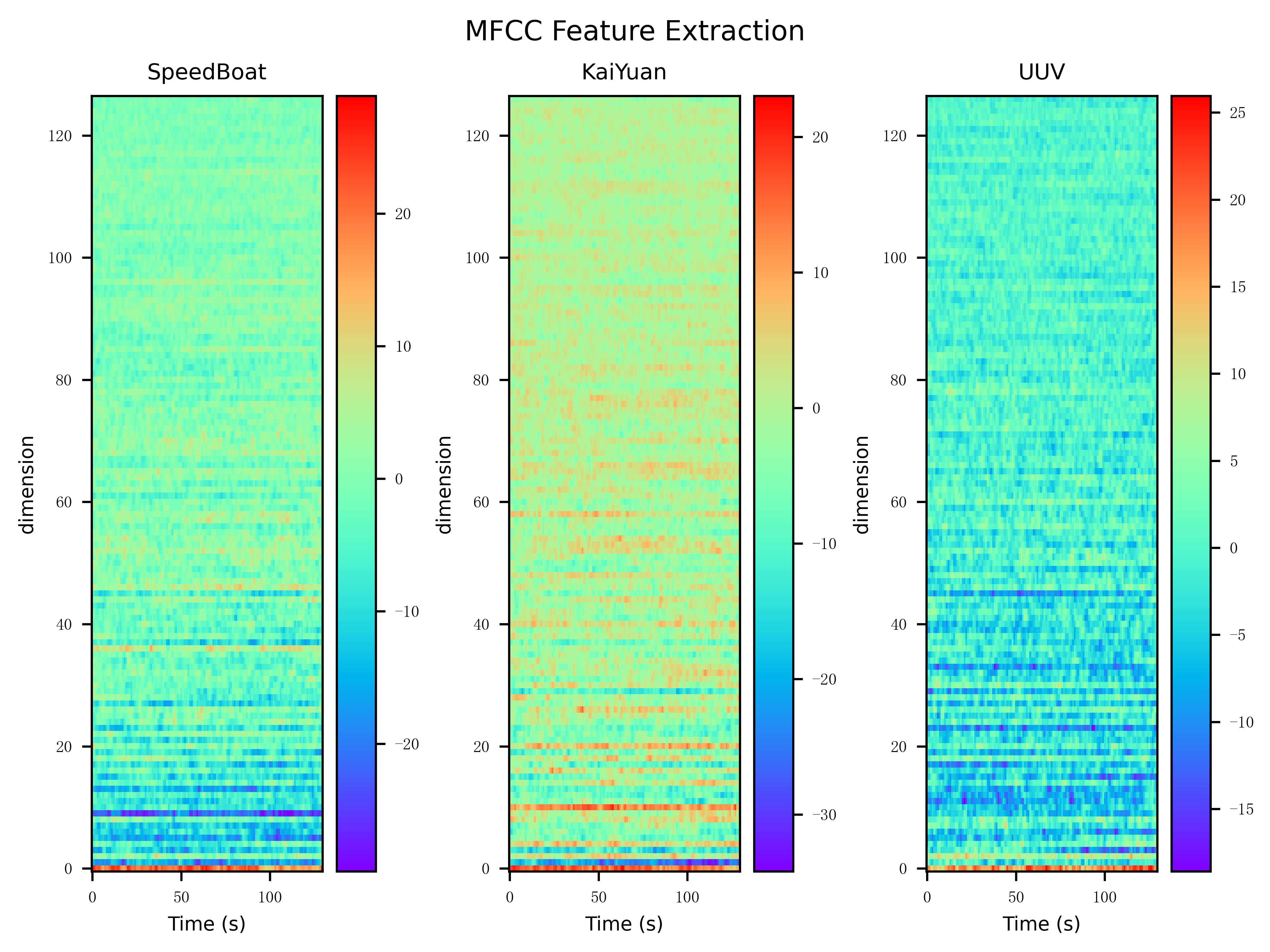}
		}
		\subfigure[PNCC ( 132*126 = 16632 )]{
			\includegraphics[width=0.21\textwidth]{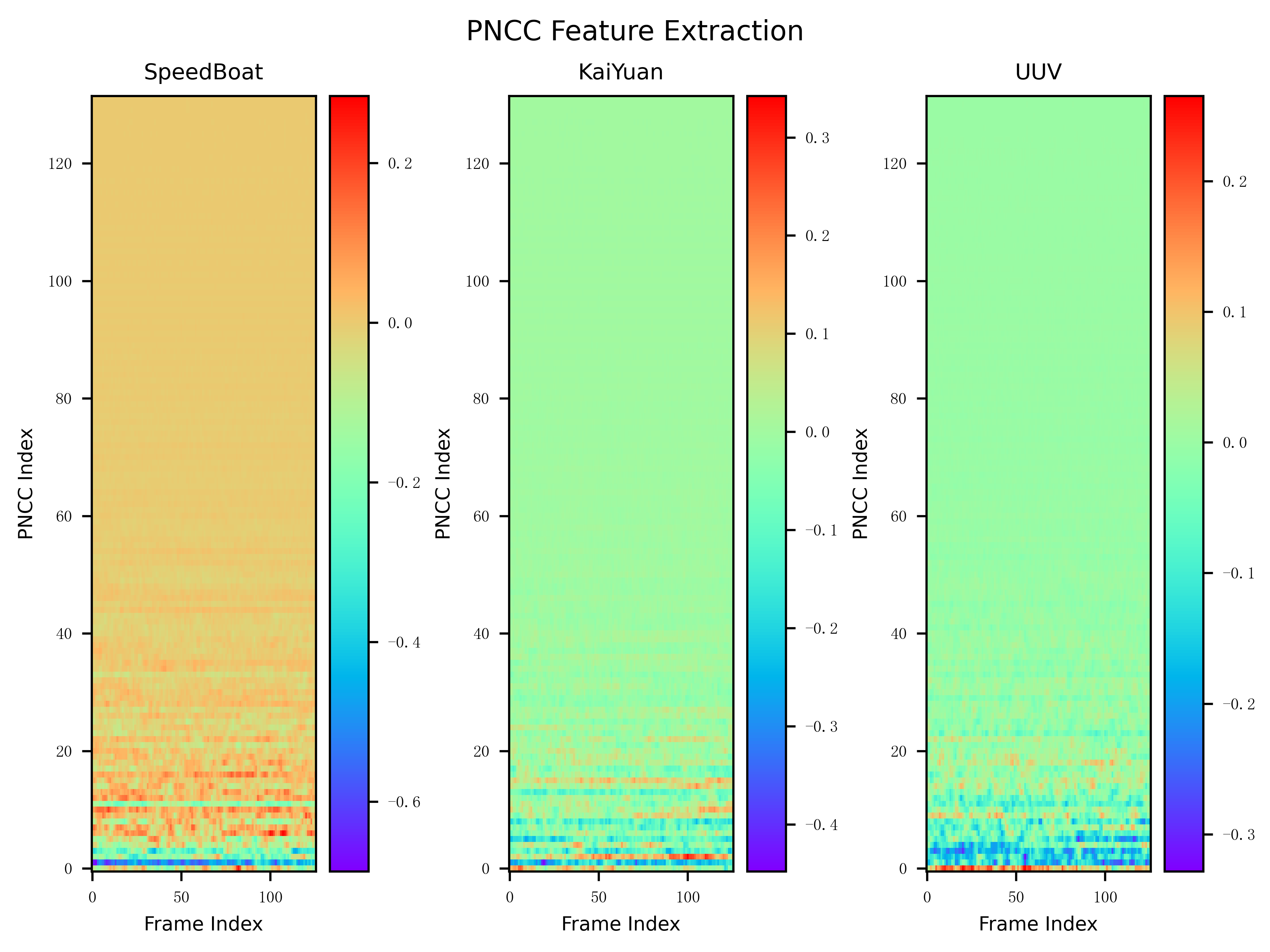}
		}
		\subfigure[GFCC ( 132*126 = 16632 )]{
			\includegraphics[width=0.21\textwidth]{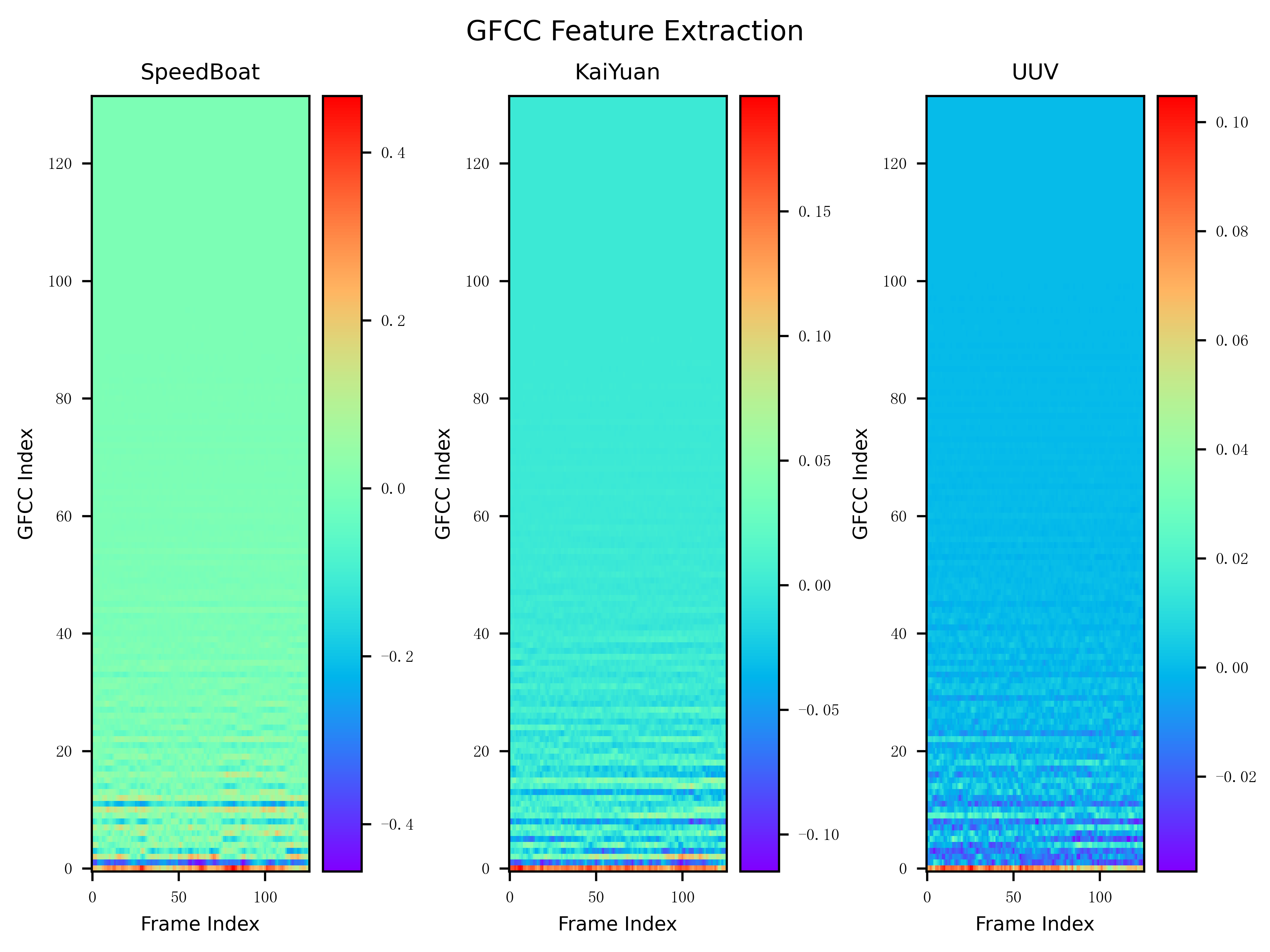}
		}
		\caption{The examples of feature extraction}
		\label{feature}
	\end{center}
\end{figure*}

\subsection{Task 2: determine if there are any targets of interest in the multi-ship mixed data}\label{subsec4}

In Task 2, our primary goal is to determine the presence of any targets of interest in the multi-ship mixed data. The selected research objects include speedboats, passenger ship KaiYuan, and underwater unmanned Vehicle UUV. Corresponding datasets were constructed for each research object, as illustrated in Table \ref{3data}. Subsequently, examples of the extracted features and their corresponding dimensions are depicted in Fig. \ref*{feature}.

To facilitate performance comparison among different features, we constrained the feature dimensions within the range of 16,384 (128128) to 16,900 (130130). Finally, deep learning networks were employed for feature classification, with the network structures consistent with Task 1. For 1D features, CNN, CRNN, and BiLSTM were exclusively used for comparative experiments. Model performance was evaluated using recall, precision, accuracy, F1 score, FNR, and FPR.

For the three target datasets established, we conducted various comparative experiments with different features and networks. To better analyze the experimental results, we summarize the results from different perspectives, forming the following three sections: Comparison of Network \ref{subsubsec1}, Comparison of Feature \ref{subsubsec2}, and Summarize \ref*{subsubsec3}.

\subsubsection{Comparison of Network}\label{subsubsec1}

In this section, to compare the performance of different networks, we calculated the average indicators for 1D and 2D features separately in experiments with different datasets using the same network. The results in the figures represent all UUV experiment outcomes with varying shades of blue bars, all KaiYuan experiment results with varying shades of green bars, and similarly, all SpeedBoat experiment outcomes with varying shades of orange bars.\\
\noindent
\small{\textbf{Comparison of network for 2D Feature Classification}}
\begin{table*}[t!]\scriptsize
	\centering
	\caption{Average results of deep learning methods for 2D feature classification(\%)}
	{\fontsize{8}{10}\selectfont
	\label{2dresulttable}
	\centering
	\begin{tabular}{cccccccc}
		\Xhline{1pt}
		experiment & net & accuracy & recall & precision & F1 score & FNR &FPR\\
		\Xhline{0.5pt}
		UUV & DenseNet & 97.78 & 96.96 & 97.93 & 97.36 & 4.34 & 0.84 \\
		UUV & ResNet & 97.56 & 96.75 & 97.59 & 97.08 & 5.43 & 1.07 \\
		UUV & CRNN2D & 96.93 & 95.45 & 97.45 & 96.30 & 8.45 & 0.59 \\
		UUV & CNN2D & 96.21 & 94.81 & 96.35 & 95.48 & 8.93 & 1.46 \\
		UUV & ECAPA-TDNN & 96.15 & 94.99 & 95.97 & 95.41 & 8.10 & 1.65\\
		KaiYuan & DenseNet & 94.52 & 93.24 & 94.39 & 93.69 & 10.51 & 3.02\\
		KaiYuan & CRNN2D & 92.73 & 91.17 &  92.32 & 91.60 & 13.36 & 4.29 \\
		SpeedBoat & DenseNet & 92.61 & 91.82 & 91.92 & 91.83 & 5.60 & 10.76 \\
		KaiYuan & ResNet & 92.49 & 90.63 & 92.17 & 91.30 & 14.77 & 3.96 \\
		KaiYuan & CNN2D & 92.19 & 90.33 & 91.93 & 90.98 & 15.10 & 4.24 \\
		SpeedBoat & ECAPA-TDNN & 92.08 & 90.78 & 91.65 & 91.14 & 4.97 & 13.48 \\
		KaiYuan & ECAPA-TDNN & 91.82 & 90.00 & 91.31 & 90.45 & 15.30 & 4.70 \\
		SpeedBoat & ResNet & 90.85 & 89.43 & 90.33 & 89.78 & 5.90 & 15.25 \\
		SpeedBoat & CNN2D & 89.14 & 87.98 & 88.21 & 88.00 & 8.23 & 15.81 \\
		SpeedBoat & CRNN2D & 89.05 & 88.22 & 88.00 & 87.94 & 9.11 & 14.22 \\
		UUV & BiLSTM & 88.48 & 84.68 & 87.68 & 85.39 & 24.47 & 5.20 \\
		SpeedBoat & BiLSTM & 88.06 & 86.22 & 87.18 & 86.57 & 7.74 & 19.80 \\
		KaiYuan & BiLSTM & 85.50 & 82.91 & 83.79 & 83.27 & 24.64 & 9.54 \\
		\Xhline{1pt}
	\end{tabular}%
	}
\end{table*}

\begin{figure*}  
	\centering
	\includegraphics[width=1\textwidth]{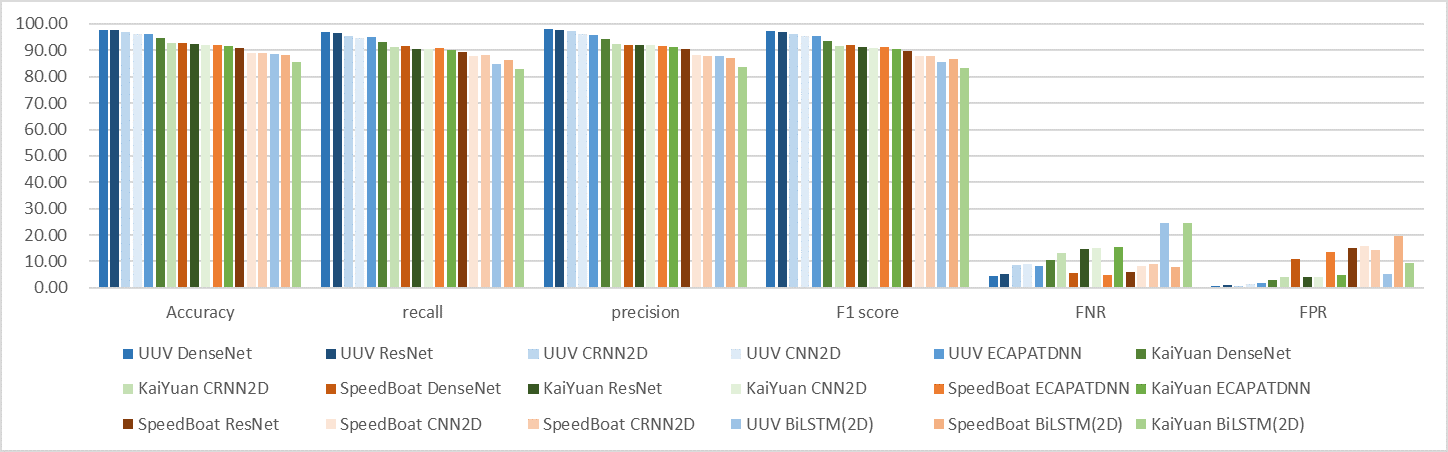}
	\caption{Average experimental results of deep learning methods for 2D feature classification}
	\label{2dresultfig}
\end{figure*}

The experimental results of different networks for 2D feature classification are shown in the Table \ref{2dresulttable} and Fig. \ref{2dresultfig}. We can observe that in the UUV experiment, the network performance from best to worst is DenseNet, ResNet, CRNN, CNN, ECAPA-TDNN, BiLSTM. In the KaiYuan experiment, the network performance from best to worst is DenseNet, CRNN, ResNet, CNN, ECAPA-TDNN, BiLSTM. In the SpeedBoat experiment, the network performance from best to worst is DenseNet, ECAPA-TDNN, ResNet, CNN, CRNN, BiLSTM. In the UUV experiment using the DenseNet network, the average recognition accuracy can reach 97.78\%, 96.96\% of the data containing UUV targets are accurately recognized, and 97.93\% of the data predicted by the network containing UUV targets are correct targets. The average F1 score also reaches 97.36\%, with only 4.34\% of the data containing UUV targets being falsely recognized, and the mispredicted UUV accounts for only 0.84\% of the non-target ship samples.

Combining the results from the three experiments, it's evident that the networks demonstrating the best overall performance for 2D feature classification are DenseNet and CRNN, followed by ResNet, which exhibits certain advantages for small-scale datasets (UUV). ECAPA-TDNN, on the other hand, presents a more pronounced advantage in classifying large-scale and complex datasets (SpeedBoat). CNN performs at an average level, while the network with the poorest classification performance is BiLSTM.

To better assess network performance, we computed the scale and computational complexity of deep learning models, primarily using two indicators: model parameters and FLOPs (Floating Point Operations Per Second). Model parameters refer to the total number of trainable parameters in a deep learning model, including weights and biases in neural networks. A higher number of parameters indicate a larger model capacity, but it also implies higher computational costs for training and inference. FLOPs represent the number of floating-point operations per second, commonly used to measure the performance of computers or computing devices. In deep learning, FLOPs are often used to describe the computational complexity of a model, indicating the total number of floating-point operations. A higher FLOPs value suggests that the model requires more computational resources. These two metrics provide comprehensive insights into the scale and computational requirements of a model.

\begin{table*}[t!]\scriptsize
	\centering
	\caption{Computational complexity and scale of deep learning models}
	{\fontsize{8}{10}\selectfont
	\label{comt}
	\begin{tabular}{ccc}
		\Xhline{1pt}
		Method & FLOPS & parameters \\
		\Xhline{0.5pt}
		CNN2D & 0.96M & 0.33M\\
		CRNN2D & 4.76M & 0.07M\\
		DenseNet & 933.09M & 7.97M\\
		ECAPA-TDNN & 169.44M & 1.53M\\
		BiLSTM & 102.24M & 0.80M\\
		ResNet & 829.26M & 13.96M\\
		\Xhline{1pt}
	\end{tabular}
}
\end{table*}

The computational experimental results, as presented in the Table \ref{comt}, have been visually represented in Fig. \ref{2dresultfig} by incorporating computational complexity into the classification outcomes. In the figure, darker colors in the bars indicate higher computational complexity. From these two perspectives, it is evident that DenseNet and ResNet exhibit higher computational complexity and, correspondingly, superior classification performance. Models with moderate computational efficiency, such as ECAPA-TDNN and BiLSTM, demonstrate relatively poorer classification results. On the other hand, CNN and CRNN, being light-weight models, achieve above-average classification performance. This further underscores the principle that there is no free lunch, and pursuing exceptional accuracy inevitably requires sacrificing computational efficiency at a certain cost.

In conclusion, for underwater acoustic target recognition tasks using 2D features, we propose the following network selection strategy:

1. For small to medium-scale datasets, the CRNN (Convolutional Recurrent Neural Network) is recommended due to its lower computational complexity and relatively good classification performance.

2. For large-scale datasets, the ECAPA-TDNN (Emphasized Channel Attention, Propagation and Aggregation in Time Delay Neural Network) network is suggested. It exhibits moderate computational complexity but demonstrates a certain advantage in handling large-scale datasets.

3. If the primary focus is on achieving extremely high classification accuracy without considering computational efficiency, the DenseNet (Densely Connected Convolutional Network) can be directly chosen. It consistently presents the best classification results across datasets of different scales.

\begin{table*}[t!]\scriptsize
	\centering
	\caption{Average results of deep learning methods for 1D feature classification(\%)}
	{\fontsize{8}{10}\selectfont
	\label{1dresulttable}
	\centering
	\begin{tabular}{cccccccc}
		\Xhline{1pt}
		experiment & net & accuracy & recall & precision & F1 score & FNR &FPR\\
		\Xhline{0.5pt}
		UUV & CNN1D & 97.22 & 96.17 & 97.36 & 96.70 & 6.62 & 1.04 \\
		UUV & CRNN1D & 96.80 & 95.72 & 96.77 & 96.18 & 7.16 & 1.39 \\
		UUV & BiLSTM & 96.50 & 94.84 & 97.08 & 95.69 & 9.59 & 0.33 \\
		SpeedBoat & CNN1D & 94.82 & 89.14 & 89.87 & 89.46 & 6.41 & 15.32 \\
		KaiYuan & CRNN1D & 93.82 & 92.52 & 93.34 & 92.87 & 11.25 & 3.70 \\
		KaiYuan & CNN1D & 92.74 & 91.94 & 91.84 & 91.82 & 10.37 & 5.74 \\
		SpeedBoat & CRNN1D & 91.53 & 90.67 & 90.72 & 90.64 & 6.51 & 12.16 \\
		KaiYuan & BiLSTM & 90.24 & 87.79 & 89.71 & 88.58 & 19.34 & 5.08 \\
		SpeedBoat & BiLSTM & 86.54 & 84.08 & 85.71 & 84.59 & 7.84 & 24.00 \\
		\Xhline{1pt}
	\end{tabular}%
	}
\end{table*}

\begin{figure*}  
	\centering
	\includegraphics[width=1\textwidth]{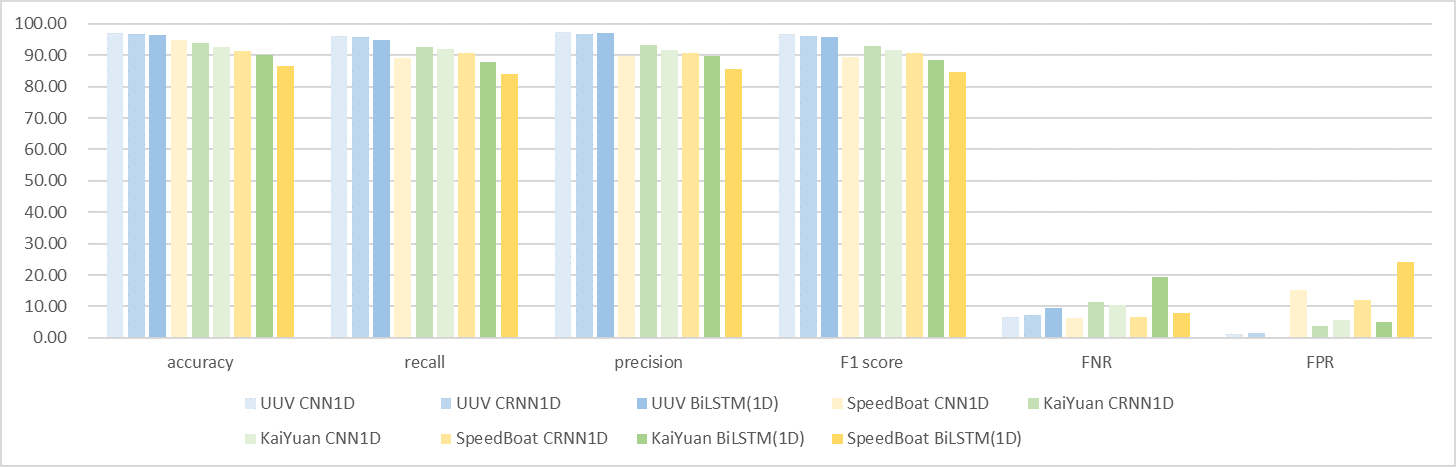}
	\caption{Average experimental results of deep learning methods for 1D feature classification}
	\label{1dresultfig}
\end{figure*}

\begin{figure*}[t!]  
	\centering
	\includegraphics[width=1\textwidth]{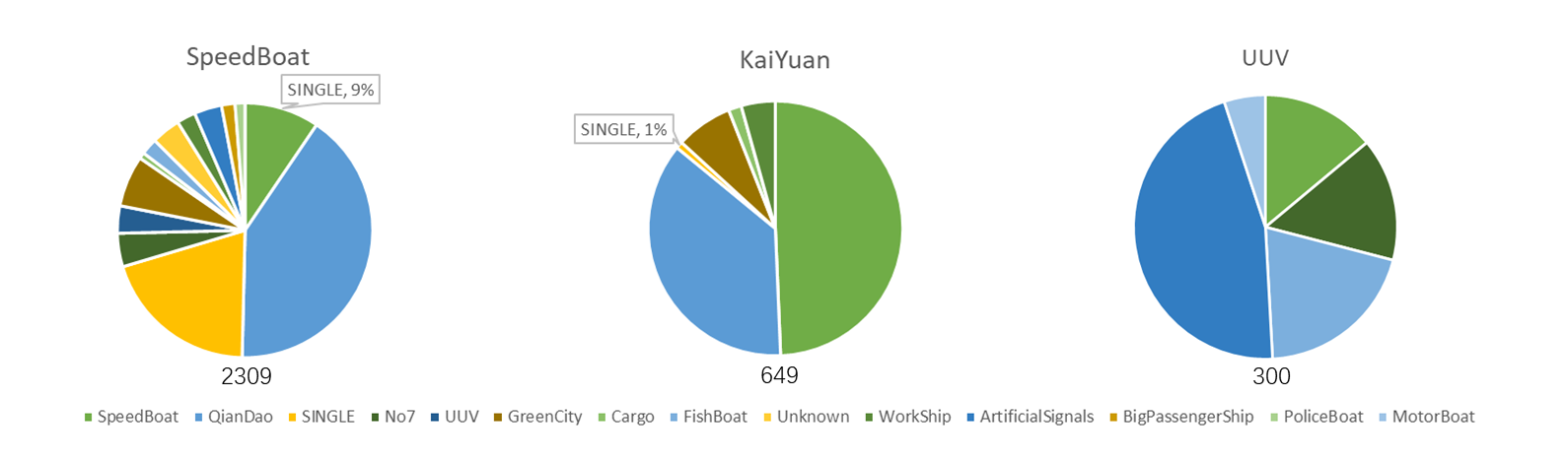}
	\caption{The data distribution for the three experiments. In the SpeedBoat experiment, the total number of vessels is 2309, while in the KaiYuan experiment, the total number of vessels is 649, and in the UUV experiment, the total number of vessels is 300.}
	\label{dis}
\end{figure*}

\begin{table}[t!]\scriptsize
	\centering
	\caption{Avearge results of the same feature (\%)}
	{\fontsize{8}{10}\selectfont
	\label{featable}
	\centering
	\begin{tabular}{ccccccc}
		\Xhline{1pt}
		feature & accuracy & recall & precision & F1 score & FNR &FPR\\
		\Xhline{0.5pt}
		Spectrum & 96.73 & 96.43 & 96.30 & 96.35 & 3.49 & 3.65 \\
		MFCC & 95.47 & 94.92 & 95.06 & 94.95 & 5.36 & 4.54 \\
		Log Mel & 94.24 & 93.24 & 93.89 & 93.48 & 7.43 & 6.09 \\
		PSD & 93.73 & 92.40 & 93.55 & 92.87 & 7.84 & 7.35 \\
		GFCC & 92.77 & 91.43 & 92.25 & 91.75 & 10.23 & 6.66 \\
		PNCC & 92.56 & 91.25 & 91.96 & 91.54 & 10.16 & 7.25 \\
		DEMON & 89.61 & 85.46 & 87.62 & 86.29 & 17.02 & 11.92 \\
		LOFAR & 87.22 & 83.72 & 86.89 & 84.83 & 20.99 & 11.58 \\
		\Xhline{1pt}
	\end{tabular}%
	}
\end{table}

\begin{table*}[t!]\scriptsize
	\centering
	\caption{Spectrum feature classification results (\%)}
	{\fontsize{8}{10}\selectfont
	\label{Fre1}
	\centering
	\begin{tabular}{>{\centering\arraybackslash}p{1.2cm}c>{\centering\arraybackslash}p{1.2cm}c>{\centering\arraybackslash}p{1.2cm}c>{\centering\arraybackslash}p{1.2cm}c>{\centering\arraybackslash}p{1.2cm}c>{\centering\arraybackslash}p{1.2cm}c>{\centering\arraybackslash}p{1.2cm}c}
		\Xhline{1pt}
		\multicolumn{1}{c}{data} & net   & accuracy   & recall & precision   & f1    & FNR & FPR \\
		\Xhline{0.5pt}
		\multicolumn{1}{c}{\multirow{3}{*}{SpeedBoat}} & \textbf{CNN1d} & \textbf{95.40} & \textbf{94.57} & \textbf{95.27} & \textbf{94.89} & \textbf{2.69} & \textbf{8.18}\\
		\cline{2-8}          & CRNN1D & 94.09 & 93.71& 93.34 & 93.52 & 5.03 & 7.55\\
		\cline{2-8}          & BiLSTM & 93.66 & 93.50 & 92.69 & 93.07 & 5.99 & 7.00\\
		\Xhline{0.5pt}
		\multicolumn{1}{c}{\multirow{3}{*}{KaiYuan}} & CNN1d & 97.11 & 96.71 & 96.74 & 96.72 & 4.44 & 2.14\\
		\cline{2-8}      & \textbf{CRNN1D} & \textbf{97.15} & \textbf{97.15} & \textbf{96.47} & \textbf{96.80} & \textbf{2.87}  & \textbf{2.84}\\
		\cline{2-8}          & BiLSTM & 95.45 & 94.82 & 94.86 & 94.84 & 7.03 & 3.34\\
		\Xhline{0.5pt}
		\multicolumn{1}{c}{\multirow{3}{*}{UUV}} & CNN1d & 98.64& 98.49 & 98.38 & 98.43 & 1.91 & 1.11\\
		\cline{2-8}          & CRNN1D & 99.49& 99.23 & 99.60 & 99.41 & 1.47 & 0.07\\
		\cline{2-8}          & \textbf{BiLSTM} & \textbf{99.56} & \textbf{99.68} & \textbf{99.32} & \textbf{99.49} & \textbf{0.00} & \textbf{0.63}\\
		\Xhline{1pt}
	\end{tabular}%
	}
\end{table*}
\noindent
\small{\textbf{Comparison of network for 1D Feature Classification}}

The network classification results for 1D data are illustrated in Table \ref{1dresulttable} and Fig. \ref{2dresultfig}. In the UUV experiment, the CNN network outperformed CRNN and BiLSTM. In the KaiYuan experiment, CRNN demonstrated superior performance to CNN based on FPR and FNR indicators; however, CNN outperformed CRNN based on the other four metrics. In the SpeedBoat experiment, CNN exhibited better accuracy compared to CRNN, whereas CRNN outperformed CNN in terms of FPR, F1 score, and recall metrics. The performance of both networks was relatively comparable in terms of FNR and precision. It's worth noting that BiLSTM consistently showed poorer performance across all experiments. With the exception of the SpeedBoat experiment based on BiLSTM, the classification results for all other networks consistently achieved above 90\%. Notably, the UUV experiment using CNN1D demonstrated an impressive average classification result of 97.22\%.

Combining the above experimental results, we can infer that 1D features contain relatively simple information, and satisfactory results can be achieved by employing straightforward network classification. Therefore, for underwater acoustic target recognition tasks based on 1D features, we propose the following strategy: For small-scale datasets (UUV), CNN networks can be used for classification, while for medium and large-scale datasets (KaiYuan, SpeedBoat), CRNN networks can be employed for classification.\\
\noindent
\small{\textbf{Comparison of network for three datasets experiments}}

Based on Fig. \ref{2dresultfig} and Fig. \ref{1dresultfig}, it can be observed that in the majority of metrics, the UUV experiment outperforms the KaiYuan experiment, while the results of the KaiYuan experiment are also superior to those of the SpeedBoat experiment. Particularly notable is the FPR metric, where the indicator FPR is significantly higher in the SpeedBoat experiment compared to the UUV and KaiYuan experiments. This is primarily due to the diverse range of vessels included in the SpeedBoat experiment, resulting in more complex combinations of different ships. We have depicted pie charts corresponding to the vessels in each of the three experimental datasets, as shown in Fig. \ref{dis}. The SpeedBoat experiment comprises a total of 2309 data points, involving 13 types of ships such as QianDao and GreenCity; the KaiYuan experiment includes 649 data points with ships like WorkShip and PoliceBoat; the UUV experiment, with only 300 data points, features vessels like PoliceBoat and FishBoat. Under the same feature and network conditions, the model exhibits better classification performance on smaller and less complex datasets, leading to the result where the UUV experiment performs better than the KaiYuan experiment, which, in turn, outperforms the SpeedBoat experiment.

Furthermore, in the evaluation of the FNR metric, we observed that the KaiYuan experiment had a higher FNR than the UUV experiment, and both were higher than the SpeedBoat experiment, contrary to the conclusions drawn earlier. We speculate that the reason of this experimental result is related to the composition of the data. Firstly, the SpeedBoat dataset contains 9\% single-target data, which is advantageous for the model to learn the characteristics of SpeedBoat targets, resulting in a low false-negative rate. For the KaiYuan dataset, although it includes 1\% single-target data, the proportion is too small. Additionally, compared to the UUV data, there is a serious imbalance in the types of samples, which may be the reason for the slightly higher false-negative rate in the KaiYuan dataset. As for the UUV dataset, although it does not contain single-target data, its sample combinations are relatively simple and evenly distributed, resulting in a moderate false-negative rate.

\subsubsection{Comparison of Feature}\label{subsubsec2}

In the feature comparison analysis, we consider that an excellent feature must be applicable in different environments and networks. Therefore, we directly average the results obtained for each feature, as shown in Table \ref{featable}. Among all features, the spectrum performs the best while MFCC features perform the best among the 2D features, with both achieving accuracies of over 95\%. Our experimental results indicate that the spectrum and MFCC features of ship-radiated noise play crucial roles in classification. The spectrum of ship-radiated noise consists of continuous, line, and modulation spectra, providing essential information about ship operations, engine status, and underwater environments. Meanwhile, MFCC features, designed based on human auditory perception, process signals through Mel filters and employ methods like logarithmic scaling and Discrete Cosine Transform (DCT) to reduce redundant features. This approach better captures features at different frequencies in ship-radiated noise.

Regarding other features, PNCC utilizes power-law nonlinearity instead of the traditional logarithmic nonlinearity in MFCC coefficients. It introduces a noise suppression algorithm based on asymmetric filtering to mask background excitation and a module for implementing temporal masking. However, these adjustments are disadvantageous for recognizing ship noise signals. GFCC features, employing Gammatone filters, enhance recognition of impulse signals. Still, as ship noise contains very few impulse signals, their recognition effect izs not as robust as that of MFCC features. DEMON spectral features respond to information such as the number of propeller blades on a ship, but since the signals identified in this paper are multi-target signals, this feature poorly distinguishes between different signals. LOFAR features are more suitable for finding narrow-band features in signals and low signal-to-noise ratio situations.

\subsubsection{Summarize}\label{subsubsec3}

Based on the raw experimental results, we observe that in the KaiYuan experiment, the best-performing recognition network is the combination of CRNN network with spectrum features. In the SpeedBoat experiment, the best-performing recognition network is the combination of CNN network with spectrum features. In the UUV experiment, the best-performing recognition network is the combination of BiLSTM network with spectrum features. Spectrum features perform excellently in all three experiments, and the complete experimental results for this feature are shown in the Table \ref*{Fre1}. The highest accuray for features containing targets in the three experiments can reach 97.15\%, 95.40\%, and 99.56\%, respectively. Moreover, they exhibit extreme stability across ten tests, with the vast majority of variances nearly approaching zero. Even with networks like BiLSTM that demonstrate comparatively poor overall performance, using Spectrum features can still yield high accuracy. This also indirectly indicates that in the field of underwater target recognition, the importance of feature decision-making may surpass that of network decision-making.

\begin{table*}[t!]\scriptsize
	\centering
	\caption{The strategies of feature and deep learning network selection for underwater acoustic target recognition tasks}
	{\fontsize{8}{10}\selectfont
	\label{select}
	\centering
	\begin{tabular}{c|c|c|c}
		\Xhline{1pt}
		& 1D feature:spectrum   & \multicolumn{2}{c}{2D feature: MFCC} \\
		\Xhline{0.5pt}
		small scale dataset & CNN & CRNN & \multirow{3}{*}{Excluding costs: DenseNet}  \\
		\cline{1-3}
		medium scale dataset & CRNN &CRNN &  \\
		\cline{1-3}
		massive scale dataset & CRNN & ECAPA-TDNN & \\
		\Xhline{1pt}
	\end{tabular}%
		}
\end{table*}

Based on our experimental results, we propose the following strategy of feature and deep learning network selection for underwater acoustic target recognition tasks, which has been shown in Table \ref*{select}. Firstly, feature selection is more crucial than network selection. Spectrum is preferred as the choice for 1D features, while MFCC features are preferred for 2D features. Secondly, for 1D feature classification, CNN can be chosen for small-scale datasets, while CRNN is suitable for medium to large-scale datasets. For 2D feature classification, CRNN is recommended for small to medium-scale datasets, while ECAPA-TDNN is suitable for massive-scale datasets. If high recognition accuracy is desired, DenseNet network can be selected. We hope that our experimental results can provide some reference for future related research.

\section{Conclusion}
\label{sec4}
This paper introduces QiandaoEar22, an underwater acoustic multi-target mixing dataset constructed through the field experiment collection in Qiandao Lake. The data set contains 9 hours and 28 minutes of ship-radiated acoustic noise and 21 hours and 58 minutes of background noise. To demonstrate the availability of QiandaoEar22, we executed two experimental tasks. The first task focuses on assessing the presence of ship-radiated noise, while the second task involves identifying specific ships within the recognized targets in the multi-ship mixed data. For latter task, we extracted eight features from the data and employed six deep learning networks for classification, aiming to evaluate and compare their performance. We set speedboat, passenger ship KaiYuan, and underwater unmanned vehicle UUV as the research object. The best recognition accuracy achieved is 99.56\%. Additionally, based on our findings, we provide advice on selecting appropriate features and deep learning networks, which may offer valuable insights for related research. In the future, we hope to establish some form of collaborative network to better learn the semantic information contained in different features, and to develop more efficient deep learning networks with higher robustness to better accomplish underwater acoustic target recognition tasks.

\backmatter

\bmhead*{Acknowledgements}  

This work was funded jointly by Natural Science Foundation of Shanghai (Grant No. 22ZR1475700), Youth Innovation Promotion Association CAS (Grant No. 2021022), and the development fund for Shanghai talents. We also sincerely thank Prof. Weijie Xu, Prof. Weicai Zhang, and Mr. Bang Jin for their great support during the experiment.

\bmhead*{Author contributions}

Xiaoyang Du contributed to the coding, analysis of experimental data, and drafting of the original manuscript. Feng Hong contributed to the acquisition of the research project, design and execution of the experimental data collection scheme, guidance on algorithm analysis research, and provided advices on the revision of the draft manuscript. All authors have read and agreed to the manuscript.

\bmhead*{Funding}

This work was funded jointly by Natural Science Foundation of Shanghai (Grant No. 22ZR1475700), Youth Innovation Promotion Association CAS (Grant No. 2021022), and Young Talent Cultivation Program of Shanghai Branch of CAS.

\bmhead*{Availability of data and materials}

The authors confirm that the data supporting the findings of this study are available within the article.

\section*{Declarations}

\bmhead*{Ethics approval and consent to participate}

Not applicable.

\bmhead*{Consent for publication}

All authors consent to the publication of this manuscript.

\bmhead*{Competing interest}

The authors declare that they have no known competing financial interests or personal relationships that could have appeared to influence the work reported in this paper.


\bibliography{sn-bibliography}


\begin{thebibliography}{22}
\ifx \bisbn   \undefined \def \bisbn  #1{ISBN #1}\fi
\ifx \binits  \undefined \def \binits#1{#1}\fi
\ifx \bauthor  \undefined \def \bauthor#1{#1}\fi
\ifx \batitle  \undefined \def \batitle#1{#1}\fi
\ifx \bjtitle  \undefined \def \bjtitle#1{#1}\fi
\ifx \bvolume  \undefined \def \bvolume#1{\textbf{#1}}\fi
\ifx \byear  \undefined \def \byear#1{#1}\fi
\ifx \bissue  \undefined \def \bissue#1{#1}\fi
\ifx \bfpage  \undefined \def \bfpage#1{#1}\fi
\ifx \blpage  \undefined \def \blpage #1{#1}\fi
\ifx \burl  \undefined \def \burl#1{\textsf{#1}}\fi
\ifx \doiurl  \undefined \def \doiurl#1{\url{https://doi.org/#1}}\fi
\ifx \betal  \undefined \def \betal{\textit{et al.}}\fi
\ifx \binstitute  \undefined \def \binstitute#1{#1}\fi
\ifx \binstitutionaled  \undefined \def \binstitutionaled#1{#1}\fi
\ifx \bctitle  \undefined \def \bctitle#1{#1}\fi
\ifx \beditor  \undefined \def \beditor#1{#1}\fi
\ifx \bpublisher  \undefined \def \bpublisher#1{#1}\fi
\ifx \bbtitle  \undefined \def \bbtitle#1{#1}\fi
\ifx \bedition  \undefined \def \bedition#1{#1}\fi
\ifx \bseriesno  \undefined \def \bseriesno#1{#1}\fi
\ifx \blocation  \undefined \def \blocation#1{#1}\fi
\ifx \bsertitle  \undefined \def \bsertitle#1{#1}\fi
\ifx \bsnm \undefined \def \bsnm#1{#1}\fi
\ifx \bsuffix \undefined \def \bsuffix#1{#1}\fi
\ifx \bparticle \undefined \def \bparticle#1{#1}\fi
\ifx \barticle \undefined \def \barticle#1{#1}\fi
\bibcommenthead
\ifx \bconfdate \undefined \def \bconfdate #1{#1}\fi
\ifx \botherref \undefined \def \botherref #1{#1}\fi
\ifx \url \undefined \def \url#1{\textsf{#1}}\fi
\ifx \bchapter \undefined \def \bchapter#1{#1}\fi
\ifx \bbook \undefined \def \bbook#1{#1}\fi
\ifx \bcomment \undefined \def \bcomment#1{#1}\fi
\ifx \oauthor \undefined \def \oauthor#1{#1}\fi
\ifx \citeauthoryear \undefined \def \citeauthoryear#1{#1}\fi
\ifx \endbibitem  \undefined \def \endbibitem {}\fi
\ifx \bconflocation  \undefined \def \bconflocation#1{#1}\fi
\ifx \arxivurl  \undefined \def \arxivurl#1{\textsf{#1}}\fi
\csname PreBibitemsHook\endcsname

\bibitem[\protect\citeauthoryear{D'amico and Pittenger}{2009}]{d2009brief}
\begin{botherref}
\oauthor{\bsnm{D'amico}, \binits{A.}},
\oauthor{\bsnm{Pittenger}, \binits{R.}}:
A brief history of active sonar.
Aquatic Mammals
\textbf{35}(4)
(2009)
\end{botherref}
\endbibitem

\bibitem[\protect\citeauthoryear{Zak}{2008}]{zak2008ships}
\begin{barticle}
\bauthor{\bsnm{Zak}, \binits{A.}}:
\batitle{Ships classification basing on acoustic signatures}.
\bjtitle{WSEAS Transactions on Signal Processing}
\bvolume{4}(\bissue{4}),
\bfpage{137}--\blpage{149}
(\byear{2008})
\end{barticle}
\endbibitem

\bibitem[\protect\citeauthoryear{Luo et~al.}{2023}]{luo2023survey}
\begin{barticle}
\bauthor{\bsnm{Luo}, \binits{X.}},
\bauthor{\bsnm{Chen}, \binits{L.}},
\bauthor{\bsnm{Zhou}, \binits{H.}},
\bauthor{\bsnm{Cao}, \binits{H.}}:
\batitle{A survey of underwater acoustic target recognition methods based on
  machine learning}.
\bjtitle{Journal of Marine Science and Engineering}
\bvolume{11}(\bissue{2}),
\bfpage{384}
(\byear{2023})
\end{barticle}
\endbibitem

\bibitem[\protect\citeauthoryear{Neupane and Seok}{2020}]{neupane2020review}
\begin{barticle}
\bauthor{\bsnm{Neupane}, \binits{D.}},
\bauthor{\bsnm{Seok}, \binits{J.}}:
\batitle{A review on deep learning-based approaches for automatic sonar target
  recognition}.
\bjtitle{Electronics}
\bvolume{9}(\bissue{11}),
\bfpage{1972}
(\byear{2020})
\end{barticle}
\endbibitem

\bibitem[\protect\citeauthoryear{Arveson and
  Vendittis}{2000}]{arveson2000radiated}
\begin{barticle}
\bauthor{\bsnm{Arveson}, \binits{P.T.}},
\bauthor{\bsnm{Vendittis}, \binits{D.J.}}:
\batitle{Radiated noise characteristics of a modern cargo ship}.
\bjtitle{The Journal of the Acoustical Society of America}
\bvolume{107}(\bissue{1}),
\bfpage{118}--\blpage{129}
(\byear{2000})
\end{barticle}
\endbibitem

\bibitem[\protect\citeauthoryear{Lennartsson
  et~al.}{2006}]{lennartsson2006fused}
\begin{bchapter}
\bauthor{\bsnm{Lennartsson}, \binits{R.}},
\bauthor{\bsnm{Dalberg}, \binits{E.}},
\bauthor{\bsnm{Levonen}, \binits{M.}},
\bauthor{\bsnm{Lindgren}, \binits{D.}},
\bauthor{\bsnm{Persson}, \binits{L.}}:
\bctitle{Fused classification of surface ships based on hydroacoustic and
  electromagnetic signatures}.
In: \bbtitle{OCEANS 2006-Asia Pacific},
pp. \bfpage{1}--\blpage{5}
(\byear{2006}).
\bcomment{IEEE}
\end{bchapter}
\endbibitem

\bibitem[\protect\citeauthoryear{McKenna et~al.}{2012}]{mckenna2012underwater}
\begin{barticle}
\bauthor{\bsnm{McKenna}, \binits{M.F.}},
\bauthor{\bsnm{Ross}, \binits{D.}},
\bauthor{\bsnm{Wiggins}, \binits{S.M.}},
\bauthor{\bsnm{Hildebrand}, \binits{J.A.}}:
\batitle{Underwater radiated noise from modern commercial ships}.
\bjtitle{The Journal of the Acoustical Society of America}
\bvolume{131}(\bissue{1}),
\bfpage{92}--\blpage{103}
(\byear{2012})
\end{barticle}
\endbibitem

\bibitem[\protect\citeauthoryear{Erbe}{2013}]{erbe2013underwater}
\begin{barticle}
\bauthor{\bsnm{Erbe}, \binits{C.}}:
\batitle{Underwater noise of small personal watercraft (jet skis)}.
\bjtitle{The Journal of the Acoustical Society of America}
\bvolume{133}(\bissue{4}),
\bfpage{326}--\blpage{330}
(\byear{2013})
\end{barticle}
\endbibitem

\bibitem[\protect\citeauthoryear{Roth et~al.}{2013}]{roth2013underwater}
\begin{barticle}
\bauthor{\bsnm{Roth}, \binits{E.H.}},
\bauthor{\bsnm{Schmidt}, \binits{V.}},
\bauthor{\bsnm{Hildebrand}, \binits{J.A.}},
\bauthor{\bsnm{Wiggins}, \binits{S.M.}}:
\batitle{Underwater radiated noise levels of a research icebreaker in the
  central arctic ocean}.
\bjtitle{The Journal of the Acoustical Society of America}
\bvolume{133}(\bissue{4}),
\bfpage{1971}--\blpage{1980}
(\byear{2013})
\end{barticle}
\endbibitem

\bibitem[\protect\citeauthoryear{Santos-Dom{\'\i}nguez
  et~al.}{2016}]{santos2016shipsear}
\begin{barticle}
\bauthor{\bsnm{Santos-Dom{\'\i}nguez}, \binits{D.}},
\bauthor{\bsnm{Torres-Guijarro}, \binits{S.}},
\bauthor{\bsnm{Cardenal-L{\'o}pez}, \binits{A.}},
\bauthor{\bsnm{Pena-Gimenez}, \binits{A.}}:
\batitle{Shipsear: An underwater vessel noise database}.
\bjtitle{Applied Acoustics}
\bvolume{113},
\bfpage{64}--\blpage{69}
(\byear{2016})
\end{barticle}
\endbibitem

\bibitem[\protect\citeauthoryear{Irfan et~al.}{2021}]{irfan2021deepship}
\begin{barticle}
\bauthor{\bsnm{Irfan}, \binits{M.}},
\bauthor{\bsnm{Jiangbin}, \binits{Z.}},
\bauthor{\bsnm{Ali}, \binits{S.}},
\bauthor{\bsnm{Iqbal}, \binits{M.}},
\bauthor{\bsnm{Masood}, \binits{Z.}},
\bauthor{\bsnm{Hamid}, \binits{U.}}:
\batitle{Deepship: An underwater acoustic benchmark dataset and a separable
  convolution based autoencoder for classification}.
\bjtitle{Expert Systems with Applications}
\bvolume{183},
\bfpage{115270}
(\byear{2021})
\end{barticle}
\endbibitem

\bibitem[\protect\citeauthoryear{Chung et~al.}{2011}]{chung2011demon}
\begin{botherref}
\oauthor{\bsnm{Chung}, \binits{K.W.}},
\oauthor{\bsnm{Sutin}, \binits{A.}},
\oauthor{\bsnm{Sedunov}, \binits{A.}},
\oauthor{\bsnm{Bruno}, \binits{M.}}:
Demon acoustic ship signature measurements in an urban harbor.
Advances in Acoustics and Vibration
\textbf{2011}
(2011)
\end{botherref}
\endbibitem

\bibitem[\protect\citeauthoryear{Chen et~al.}{2021}]{chen2021underwater}
\begin{barticle}
\bauthor{\bsnm{Chen}, \binits{J.}},
\bauthor{\bsnm{Han}, \binits{B.}},
\bauthor{\bsnm{Ma}, \binits{X.}},
\bauthor{\bsnm{Zhang}, \binits{J.}}:
\batitle{Underwater target recognition based on multi-decision lofar spectrum
  enhancement: A deep-learning approach}.
\bjtitle{Future Internet}
\bvolume{13}(\bissue{10}),
\bfpage{265}
(\byear{2021})
\end{barticle}
\endbibitem

\bibitem[\protect\citeauthoryear{Zhang et~al.}{2016}]{zhang2016feature}
\begin{botherref}
\oauthor{\bsnm{Zhang}, \binits{L.}},
\oauthor{\bsnm{Wu}, \binits{D.}},
\oauthor{\bsnm{Han}, \binits{X.}},
\oauthor{\bsnm{Zhu}, \binits{Z.}}, et al.:
Feature extraction of underwater target signal using mel frequency cepstrum
  coefficients based on acoustic vector sensor.
Journal of Sensors
\textbf{2016}
(2016)
\end{botherref}
\endbibitem

\bibitem[\protect\citeauthoryear{Kim and Stern}{2016}]{kim2016power}
\begin{barticle}
\bauthor{\bsnm{Kim}, \binits{C.}},
\bauthor{\bsnm{Stern}, \binits{R.M.}}:
\batitle{Power-normalized cepstral coefficients (pncc) for robust speech
  recognition}.
\bjtitle{IEEE/ACM Transactions on audio, speech, and language processing}
\bvolume{24}(\bissue{7}),
\bfpage{1315}--\blpage{1329}
(\byear{2016})
\end{barticle}
\endbibitem

\bibitem[\protect\citeauthoryear{Zhao and Wang}{2013}]{zhao2013analyzing}
\begin{bchapter}
\bauthor{\bsnm{Zhao}, \binits{X.}},
\bauthor{\bsnm{Wang}, \binits{D.}}:
\bctitle{Analyzing noise robustness of mfcc and gfcc features in speaker
  identification}.
In: \bbtitle{2013 IEEE International Conference on Acoustics, Speech and Signal
  Processing},
pp. \bfpage{7204}--\blpage{7208}
(\byear{2013}).
\bcomment{IEEE}
\end{bchapter}
\endbibitem

\bibitem[\protect\citeauthoryear{Cheng et~al.}{}]{Cheng0Research}
\begin{botherref}
\oauthor{\bsnm{Cheng}, \binits{J.}},
\oauthor{\bsnm{Xuanming}, \binits{D.}},
\oauthor{\bsnm{Sai}, \binits{Z.}}:
Research on extraction and recognition technique to auditory features of
  underwater target based on deep learning.
In: Shanghai 2018 Marine Electronic Equipment Research Institute
\end{botherref}
\endbibitem

\bibitem[\protect\citeauthoryear{Fu et~al.}{2017}]{fu2017crnn}
\begin{bchapter}
\bauthor{\bsnm{Fu}, \binits{X.}},
\bauthor{\bsnm{Ch'ng}, \binits{E.}},
\bauthor{\bsnm{Aickelin}, \binits{U.}},
\bauthor{\bsnm{See}, \binits{S.}}:
\bctitle{Crnn: a joint neural network for redundancy detection}.
In: \bbtitle{2017 IEEE International Conference on Smart Computing
  (SMARTCOMP)},
pp. \bfpage{1}--\blpage{8}
(\byear{2017}).
\bcomment{IEEE}
\end{bchapter}
\endbibitem

\bibitem[\protect\citeauthoryear{Zhang et~al.}{2020}]{Zhang2020Classification}
\begin{barticle}
\bauthor{\bsnm{Zhang}, \binits{S.}},
\bauthor{\bsnm{Wang}, \binits{C.}},
\bauthor{\bsnm{Sun}, \binits{Q.}}:
\batitle{Classification technique for hydroacoustic target noise recognition
  based on multi-category feature fusion}.
\bjtitle{Journal of Northwestern Polytechnical University}
\bvolume{38}(\bissue{2}),
\bfpage{366}--\blpage{376}
(\byear{2020})
\end{barticle}
\endbibitem

\bibitem[\protect\citeauthoryear{Yu et~al.}{2021}]{yu2021additive}
\begin{barticle}
\bauthor{\bsnm{Yu}, \binits{D.}},
\bauthor{\bsnm{Yang}, \binits{J.}},
\bauthor{\bsnm{Zhang}, \binits{Y.}},
\bauthor{\bsnm{Yu}, \binits{S.}}:
\batitle{Additive densenet: Dense connections based on simple addition
  operations}.
\bjtitle{Journal of Intelligent \& Fuzzy Systems}
\bvolume{40}(\bissue{3}),
\bfpage{5015}--\blpage{5025}
(\byear{2021})
\end{barticle}
\endbibitem

\bibitem[\protect\citeauthoryear{He et~al.}{2016}]{he2016deep}
\begin{bchapter}
\bauthor{\bsnm{He}, \binits{K.}},
\bauthor{\bsnm{Zhang}, \binits{X.}},
\bauthor{\bsnm{Ren}, \binits{S.}},
\bauthor{\bsnm{Sun}, \binits{J.}}:
\bctitle{Deep residual learning for image recognition}.
In: \bbtitle{Proceedings of the IEEE Conference on Computer Vision and Pattern
  Recognition},
pp. \bfpage{770}--\blpage{778}
(\byear{2016})
\end{bchapter}
\endbibitem

\bibitem[\protect\citeauthoryear{Desplanques
  et~al.}{2020}]{desplanques2020ecapa}
\begin{botherref}
\oauthor{\bsnm{Desplanques}, \binits{B.}},
\oauthor{\bsnm{Thienpondt}, \binits{J.}},
\oauthor{\bsnm{Demuynck}, \binits{K.}}:
Ecapa-tdnn: Emphasized channel attention, propagation and aggregation in tdnn
  based speaker verification.
arXiv preprint arXiv:2005.07143
(2020)
\end{botherref}
\endbibitem

\end{thebibliography}

\end{document}